\documentclass[aps,prl,twocolumn,footinbib,superscriptaddress,amsmath,amsfonts,amssymb,floatfix,notitlepage]{revtex4-2} 

\usepackage{CJK}
\usepackage{physics}
\usepackage{soul}
\usepackage[caption=false,subrefformat=parens,labelformat=parens]{subfig}
\usepackage{color} 
\usepackage[table,dvipsnames]{xcolor}
\usepackage[colorlinks,linktocpage,bookmarks=false]{hyperref}
\usepackage[normalem]{ulem} %
\usepackage{wasysym} %
\usepackage{graphicx}
\usepackage{multirow} 
\usepackage{booktabs} %

\newcommand\myshade{85}
\newcommand{\prlsection}[1]{\noindent\textbf{{#1} ---}}

\newcommand{\beq}{\begin{equation}}
\newcommand{\eeq}{\end{equation}}  \newcommand{\Jring}{g}  
\newcommand{\CZO}{Ce$_2$Zr$_2$O$_7$}%
\newcommand{\CSO}{Ce$_2$Sn$_2$O$_7$}%
\newcommand{\CHO}{Ce$_2$Hf$_2$O$_7$}%

\renewcommand\[{\begin{equation}}
\renewcommand\]{\end{equation}}

\definecolor{myrulecolor}{RGB}{150,20,0}%
\colorlet{mylinkcolor}{violet}
\colorlet{mycitecolor}{YellowOrange}
\colorlet{myurlcolor}{Aquamarine}
\hypersetup{
	linkcolor  = myurlcolor!\myshade!black,
	citecolor  = mycitecolor!\myshade!black,
	urlcolor   =  myrulecolor!\myshade!black,
}

\graphicspath{{/}}

\AtBeginDocument{
	\heavyrulewidth=.08em
	\lightrulewidth=.05em
	\cmidrulewidth=.03em
	\belowrulesep=.65ex
	\belowbottomsep=0pt
	\aboverulesep=.4ex
	\abovetopsep=0pt
	\cmidrulesep=\doublerulesep
	\cmidrulekern=.5em
	\defaultaddspace=.5em
}%

\makeindex

\newcommand{\thistitle}{Experimentally tunable QED in dipolar-octupolar quantum spin ice}
\newcommand{\hxxz}{\mathcal{H}_{\rm XX}}
\newcommand{\hz}{\mathcal{H}_{\rm Z}}

\newcommand{\as}[1]{#1}%

\usepackage{comment}

\begin{document} 
\begin{CJK*}{UTF8}{gbsn} %

\title{\thistitle}

\author{Alaric Sanders}
\affiliation{TCM Group, Cavendish Laboratory, University of Cambridge, Cambridge CB3 0HE, United Kingdom}
\author{Han Yan (闫寒)} 
\affiliation{Institute for Solid State Physics, University of Tokyo, Kashiwa, 277-8581 Chiba, Japan}
\affiliation{Department of Physics and Astronomy, Rice University, Houston, TX 77005, USA}
\affiliation{Smalley-Curl Institute, Rice University, Houston, TX 77005, USA}
\author{Claudio Castelnovo}
\affiliation{TCM Group, Cavendish Laboratory, University of Cambridge, Cambridge CB3 0HE, United Kingdom}
\author{Andriy H. Nevidomskyy}  
\affiliation{Department of Physics and Astronomy, Rice University, Houston, TX 77005, USA}
\affiliation{Rice Center for Quantum Materials, Rice University, Houston, TX 77005, USA}

\date{\today}
\begin{abstract}
We propose a readily achievable experimental setting where an external magnetic field is used to tune the emergent quantum electrodynamics (eQED) of dipolar-octupolar quantum spin ice (DO-QSI). 
In $U(1)_\pi$ DO-QSI -- the proposed ground state of QSI candidates {\CZO}, {\CSO} and {\CHO} -- we show that the field can be used to control the emergent speed of light (and, consequently, the emergent fine structure constant). Depending on the field's alignment with the crystal, one may induce different speeds for the two polarizations of the emergent photons, in a fascinating analogue of the electro-optic Kerr effect.
In $U(1)_0$ DO-QSI -- yet to be uncovered experimentally -- we find a number of unusual field-induced transitions, including a transition between $0$- and $\pi$-flux QSI phases, as well as phases with frustrated flux configurations. 
We discuss experimental signatures of these effects in the spinon excitation spectrum, which can be readily accessed for instance in inelastic neutron scattering measurements. 
Our proposal opens the gate to a plethora of experimentally accessible, engineerable eQED phenomena in the emergent universes of quantum spin ice.
\end{abstract}
\maketitle
\end{CJK*} 

\prlsection{Introduction}
Quantum spin liquids (QSL) -- exotic phases of matter where strong quantum fluctuations prevent long-range order down to absolute zero temperature -- are some of the most exquisitely complex and rich behaviours observed in quantum many-body physics~\cite{moessnerTopologicalPhasesMatter2021}. Recent years have witnessed an impressive experimental effort that brought to the fore several candidate materials. While definitive experimental confirmation of a QSL state has remained elusive thus far, concerted theoretical and experimental progress has brought us ever so close to their detection. 

Gauge field theory has emerged as a fundamental unifying principle to describe spin liquid phases. A case in point is quantum spin ice (QSI)~\cite{gingrasQuantumSpinIce2014,udagawaSpinIce2021}, whose behaviour is most naturally understood in a $3+1 D$ compact $U(1)$ emergent quantum electrodynamics (eQED) picture~\cite{Hermele_Fisher_Balents} featuring gapless photons, gapped electrically charged matter (spinons) and Dirac monopoles (visons) as quasiparticles. 
Various theoretical and numerical studies~\cite{huangDynamicsTopologicalExcitations2018,shannonQuantumIceQuantum2012,udagawaSpectrumItinerantFractional2019,hosoiUncoveringFootprintsDipolarOctupolar2022} have inspected different aspects of QSI, including photons and their speed of light~\cite{BentonPhysRevB,paceEmergentFineStructure2021}, spinon dynamics~\cite{chenDistinguishingThermodynamicsSpectroscopy2023,desrochersSpectroscopicSignaturesFractionalization2023,leeGenericQuantumSpin2012,savaryCoulombicQuantumLiquids2012}, Cerenkov radiation~\cite{morampudi2020}, and vison physics~\cite{szaboSeeingLightVison2019a,kwasigrochVisongeneratedPhotonMass2020,sandersVisonCrystalQuantum2023,chenDiracMagneticMonopole2017,bergmanOrderingFrustratedPyrochlore2006}%
, many of which go far beyond the QED accessible in fundamental particle physics.

The recent theoretical prediction~\cite{HuangPhysRevLett} of dipolar-octupolar quantum spin ice (DO-QSI) and the experimental discovery of candidate materials {\CZO}~\cite{gao2024emergent,CZO.Gao,CZO.Gaulin}, {\CHO}~\cite{poreeCrystalfieldStatesDefect2022,poree2023dipolaroctupolar,bhardwaj2024arXiv}, and {\CSO}~\cite{Sibille2020NatPhys,poree2023fractional} has further expanded the playground for QSI physics. 
In these systems, the ground state doublets of the crystalline electric fields of the magnetic ions are dominated by the $\ket{J=5/2, m_J = \pm  3/2}$ wave-functions. As a consequence, the effective pseudospin-$1/2$ operators $S_i^\alpha$ ($\alpha\in\{x,y,z\}$) 
do not transform as conventional dipole moments under lattice and time-reversal symmetries~\cite{HuangPhysRevLett}, leading to important consequences for their spectroscopic signatures~\cite{liSymmetryEnrichedTopological2017,yaoPyrochloreSpinLiquid2020,chenDistinguishingThermodynamicsSpectroscopy2023}. 
Most notably, the octupolar nature of the $S^y_i$ operators
means that they do not couple, to linear order, to an applied magnetic field.

In this Letter, we show how the properties of the DO-QSI offer a unique opportunity to tune the parameters of the emergent QED via straightforward application of external magnetic fields. \as{In the matter sector, an applied field dramatically alters the spinon dispersion, while in the gauge sector, the field changes the refractive index of the photons -- in essence, an eQED analogue of the electro-optic Kerr effect.}
This process also tunes the speed of emergent photons and therefore the fine-structure constant. 
The effect is also generically anisotropic, resulting in birefringence of the emergent photons, and is directly applicable to {\CZO}, {\CSO} and {\CHO}, which are believed to reside in the $\pi$-flux QSI phase. For systems in the $0$-flux ($U(1)_0$) QSI phase, we further find that the application of external magnetic fields results in novel phase transitions into the $\pi$-flux ($U(1)_\pi$) and various mixed-flux phases. 

\as{Crucially, a region of the DO-QSI parameter space exists with dominant exchange couplings in the \textit{octupolar} channel. The resulting octupolar ice rule is not easily polarised by an applied magnetic field,
enabling control of the quantum dynamics without polarising the system.} 
The effect we predict is achievable using readily available sub-Tesla magnetic fields, in contrast to the previously proposed electric field tuning that requires very strong field ($\sim 10^4$~V/mm) for similar effects to be observed~\cite{lantagne-hurtubiseElectricFieldControl2017, mandalElectricFieldResponse2019}. 

The ability to precisely tune the dynamics of eQED using modest 
magnetic fields is an important experimental handle that gives us access to a vast range of eQED dynamics and phases 
without the need for challenging (oftentimes impossible) fine tuning of material parameters. 

\begin{figure}[th!]
\includegraphics[width=\columnwidth]{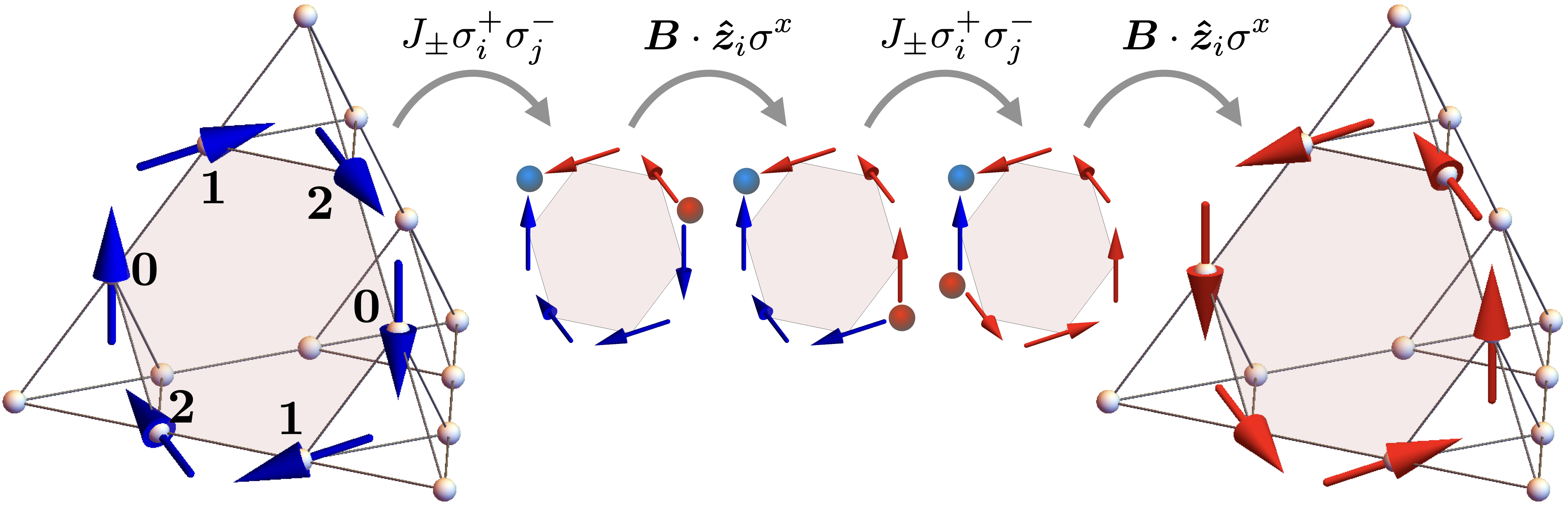}
\caption{
\label{Fig_lattice}
Schematic illustration of the pyrochlore lattice, and example of the perturbative generation of the hexagon-flip term, via both the spin-spin interaction and spin-field coupling.
Pertinent to DO-QSI candidate materials, the arrows represent the local $S^y=\sigma^z$ components of the spins.
}
\end{figure}

\prlsection{The model}
We consider DO-QSI on the pyrochlore lattice (Fig.~\ref{Fig_lattice}), described by a Hamiltonian with nearest neighbor interactions and coupling to magnetic field~\cite{PatriPhysRevResearch}
\[
\begin{split}
\mathcal{H} =& 
\sum_{\langle i j \rangle} 
\left[
  \sum_{\alpha = x, y, z} J_\alpha S_i^\alpha S_j^\alpha 
  + J_{xz} (S_i^xS_j^z +  S_i^zS_j^x)
\right] 
\\
&- \sum_i (S_i^z \tilde{g}_{zz} + S_i^x \tilde{g}_{xz}) \vb*{\hat{z}}_i \cdot \vb*{B}
\, . 
\end{split}
\]
Here, the spin operators are in the local basis, and $\vb*{\hat{z}}_i$ is the unit vector of the local $z$-axis pointing towards/away from the tetrahedron center.
An important feature of this Hamiltonian, as mentioned earlier, is that the $S^y_i$ operators do not couple to the external magnetic field to linear order, due to their octupolar symmetry~\cite{HuangPhysRevLett}. 

We study the effect of tuning the magnetic field $\vb*{B}$ in DO-QSI with dominant octupolar interaction $J_y$. 
For simplicity, we take $J_\pm\equiv J_z = J_x \ll J_y$, $J_{xz} = 0$, and $\tilde{g}_{xz} = 0$, $\tilde{g}_{zz} = 1$. 
The Hamiltonian is more conveniently written as a dominant part plus a perturbation, $\mathcal{H} = \mathcal{H}_I  + \mathcal{H}'$, where 
\begin{align}
\label{eq:H'}
\mathcal{H}_I & = 
\sum_{\langle i j \rangle} J_y \sigma_i^z \sigma_j^z, \\
\mathcal{H}' &= 
\sum_{\langle i j \rangle} 
\frac{J_\pm}{2} (\sigma_i^+\sigma_j^- + \sigma_i^-\sigma_j^+ )
-\sum_i \frac{\vb*{\hat{z}}_i \cdot \vb*{B}}{2}
(\sigma_i^+ + \sigma_i^-) 
\, . 
\nonumber
\end{align}
Here, we relabeled $S^y = \sigma_i^z$, $S^z \pm i S^x = \sigma^\pm_i$, to bring the Hamiltonian to a more customary notation~\cite{Hermele_Fisher_Balents}. 

The low energy Hilbert space is approximately spanned by the ice states, i.e., product states that satisfy the 2-in-2-out rules in the $\sigma_i^z$ basis, which are ground states of $\mathcal{H}_I$. 
In the presence of the non-commuting $\mathcal{H}'$, the ground states and low-energy excitations (i.e., photons and visons of the eQED) are coherent quantum superpositions of the ice states. 

The aforementioned absence of coupling to the $\sigma^z_i=S^y_i$ pseudo-spin components means that applied magnetic field action is purely transverse, \textit{without} biasing a preferred polarisation of the ice states. 
One therefore expects the field to tune the DO-QSI phase  but not altogether disrupt it 
up to $|\vb*{B}| \sim J^y$ (i.e., well in excess of the one Tesla scale)~\cite{liSymmetryEnrichedTopological2017}. 

The low-energy dynamics are determined by matrix elements between ice states generated perturbatively. Note that individual terms in $\mathcal{H}'$ take the system out of the ice state manifold by creating pairs of spinons. 
It is only at higher orders in perturbation theory in $\mathcal{H}'$ -- creating, moving and annihilating spinons around closed, non-self-retracing paths -- that one finds quantum processes connecting different ice states.
The simplest such process takes place around an elementary hexagonal loop on the lattice, illustrated in Fig.~\ref{Fig_lattice} and embodied by 
Hamiltonian terms 
of the form 
\begin{equation}
\mathcal{H}_\text{eff}
= \sum_{p \in \{\text{all } \hexagon \}} 
\Jring_p 
\left[ 
  \mathcal{O}_{p} + \mathcal{O}_{p}^\dagger 
\right]
+ \ldots 
\, , 
\label{eq:eff_hamiltonian}
\end{equation} 
where $\mathcal{O}_{p} = \sigma_1^+\sigma_2^-\sigma_3^+\sigma_4^-\sigma_5^+\sigma_6^-$ is the hexagon-flip operator, and $\Jring_p$ is a function of $J_\pm$ and $\vb*{B}$. 

There are four types of hexagons facing different directions.  
They generally have different $\Jring_{p}$ values due to the non-uniform magnetic field coupling proportional to $\vb*{\hat{z}}_i \cdot \vb*{B}$ on different sublattice sites. 
To leading order, we find~\cite{SM}
\[
\label{EQN_general_Jring}
\Jring_{p} = 
\frac{3 J_\pm^3}{2J_y^2}
+ 
\frac{5 J_\pm^2}{4 J_y^3} 
( \vb*{\hat{n}}_p \cdot \vb*{B} )^2 
\, ,
\] 
where $\vb*{\hat{n}}_p$ is the unit vector normal to the hexagon $p$. 

We will often discuss the four hexagons surrounding the 4-tetrahedron object shown in Fig.~\ref{Fig_lattice}.
Notice that each hexagon contains $3$ of the $4$ sublattice sites; therefore we can conveniently label the hexagon types by the sublattice index $\mu \in \{0,1,2,3\}$ that they do not contain. 
For example, Fig.~\ref{Fig_lattice} highlights a hexagon containing sublattice sites $0,1,2$, and we label it being of type $\mu = 3$. 

\begin{figure}[t!]
\includegraphics[width=\columnwidth]{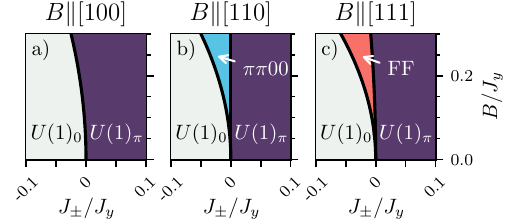}
\caption{
\label{Fig_Jring_all}
Panels (a-c): $B-J_\pm$ phase diagrams for [100], [110] and [111] applied fields. Zero-flux $U(1)_0$ and $\pi$-flux $U(1)_\pi$ states compete with the mixed-flux $\pi\pi 00$ and a continuously-varying frustrated flux phase, see Fig.~\ref{fig:magnetic_field_tuning} for further details. Phase transitions are shown as solid black lines. 
}
\end{figure}

In order to understand the role of this coupling constant in the Hamiltonian $\mathcal{H}_{\text {eff}}$ in Eq.~\eqref{eq:eff_hamiltonian}, it is convenient to use the Villain expansion~\cite{villainInsulatingSpinGlasses1979},
$
\sigma^{+}_j = e^{ia_j/2}\sqrt{(S+1/2)^{2}- (\sigma_j^z)^2} \; e^{ia_j/2}
$,
which recreates the spin algebra using the $U(1)$ operators $a_i$ canonically conjugate to $\sigma^z_i$, $[a_i, \sigma_j^z] = i \delta_{ij}$. If we take the limit $S\to \infty$, we recognize that $\mathcal{O}_p \simeq (S+1/2)^6 e^{i \Phi_p}$, to leading order, where $\Phi_p = \nabla_p\times a$ are the gauge fluxes of the emergent QED magnetic field through the hexagon~\cite{Hermele_Fisher_Balents,kwasigrochSemiclassicalApproachQuantum2017}.
Eq.~\eqref{eq:eff_hamiltonian} thus becomes $\mathcal{H}_\text{eff} \propto \sum_\mu g_\mu \cos(\Phi_\mu)$.
This highlights the importance of the sign of $J_\pm$ in Eq.~\eqref{EQN_general_Jring}. If $J_\pm>0$, we have $\Jring>0$ regardless of $\vert\vb*{B}\vert$, and the $U(1)_\pi$ state with $\Phi_p \equiv \pi$ remains always lowest in energy. 
If $J_\pm < 0$, however, a competition arises between $J_\pm$ and $\vert\vb*{B}\vert$ capable of changing the sign of $\Jring$ independently on the four sublattices, depending on the direction of the applied field. 
In this case, the ground state flux configuration must be determined by minimizing the semiclassical energy $2 \sum_{\mu = 0}^3 g_\mu \cos(\Phi_\mu)$ over $\Phi_\mu \in [-\pi,\pi]^4$ subjected to the quantization condition $\sum_{\mu=0}^3\Phi_\mu = 0$ ($\mathrm{mod}\; 2\pi$) around the four-tetrahedron structure shown in Fig.~\ref{Fig_lattice}~\cite{sandersVisonCrystalQuantum2023,szaboSeeingLightVison2019a}. We plot the resulting phase diagram in Fig.~\ref{Fig_Jring_all}. 
This procedure yields four possible ground states: the standard $U(1)_\pi$ and $U(1)_0$ states, in addition to a mixed-flux $\pi\pi 0 0$ state (in which precisely two hexagon sublattices carry flux $\pi$, and two have vanishing flux); and a continuously variable, non-quantized flux arrangement which we dub `frustrated flux' (FF). 
Photon dynamics correspond to quadratic fluctuations about these ground states.

\begin{figure}[t!]
    \includegraphics[width=\columnwidth]{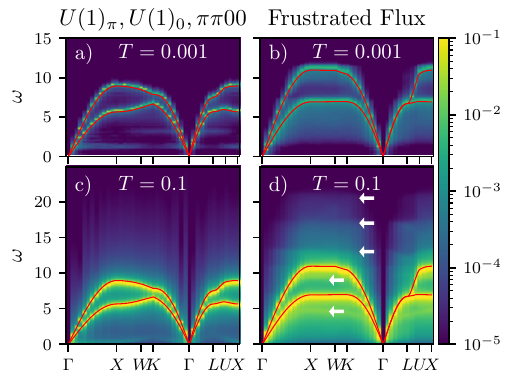}
    \caption{
    \label{fig:analytics+numerics}
    Numerical (color map) and analytical (red traces) dynamical $S^yS^y$ correlations in the $\pi\pi 00$, $U(1)_0$, $U(1)_\pi$, $\Jring_\mu = ( \pm 1/2, \pm 1/2, [\pm]1, [\pm]1)$ phases (left column); and in the frustrated-flux, $\Jring_\mu = (-2, 1, 1, 1)$ phase (right column). As discussed in the main text, the photon dispersion is not affected by changing pairs of signs of $g$, hence the identical correlations in all three phases in the left panels.
    White arrows in panel d) mark shoulders in the spectral weight due to three-photon up-conversion processes, which do not occur in the non-frustrated phases on the left. Both $T$ and $\Jring_\mu$ are in units of $J_\pm^3/J_y^2$.
    }
\end{figure}

\prlsection{Photon dynamics: emergent Kerr effect}
We calculated the spectrum of the emergent photon by two methods: (i) we performed a semiclassical path integral simulation using the method of Ref.~\onlinecite{szaboSeeingLightVison2019a}; and (ii) we 
used
the large-$S$ formalism of Refs.~\onlinecite{BentonPhysRevB,kwasigrochSemiclassicalApproachQuantum2017, kwasigrochVisongeneratedPhotonMass2020}.
The details of these calculations may be found in the Supplemental Material~\cite{SM}.
We see from Fig.~\ref{fig:analytics+numerics} that the strongest features in the simulations are precisely the modes of the quadratic theory. 

First, we remark on the rich physics achievable in the $J_\pm > 0$ case relevant to the known cerium pyrochlore materials.
Generically, magnetic field couples differently to the four types of hexagons, thus breaking the crystalline symmetry, to which quantum water ice has similar phoenomenon~\cite{Benton2016PhysRev}.
As a result, the degeneracy of the photon branches is lifted or, in the language of optics, the medium becomes birefringent, as seen in Fig.~\ref{fig:analytics+numerics}. Near the $\Gamma$ point, the splitting -- and therefore the change in speed of light $c$ of eQED for different polarizations -- scales as $\Delta c \propto  B^2 /J_y$, in a fascinating emergent analogue of the electro-optical Kerr effect (except with the magnetic, rather than electric field being applied here). For instance, using the parameters of {\CHO} given in Ref.~\onlinecite{poree2023dipolaroctupolar}, $\Delta c / c \sim 1$ can be achieved at the easily accessible field of $\sim 0.3$~T. 

In the special case of $\vb*{B} \parallel$~[100], all hexagon types have the same value of $\Jring_\mu$: $\Jring_\mu \propto (1, 1, 1, 1)$.
In this case the speed of light is altered without splitting the photon modes.
This in turn affects the fine structure constant $\alpha=q^2/(\hbar c)$ ($q$ being the spinon charge), illustrating the fascinating level of control achievable with even modest magnetic fields.

\begin{figure}[t!]
    \includegraphics[width=\columnwidth]{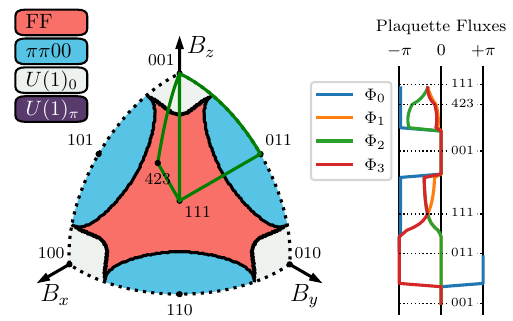}
    \caption{
    \label{fig:magnetic_field_tuning}
    Field direction dependence of the ground state flux phase in the interesting case $\| \vb*{B} \| = \sqrt{4.2 J_y J_\pm}$, where the effect of the magnetic field is comparable to the intrinsic hexagon-flip terms. Left panel: Phase diagram at constant magnetic field, labeled by high-symmetry directions on the unit sphere. Right panel: variation of the four fluxes as the magnetic field is rotated along the green path in the left panel. 
    }
\end{figure}

\vspace{1mm}
\prlsection{Tuning eQED phases with applied fields}
We next consider the case $J_\pm < 0$, in which the QSI is in the $U(1)_0$ phase in the absence of magnetic field. In addition to the birefringent physics discussed above, the applied magnetic field now also tunes the sign of $\Jring_p$, allowing one to change the ground state flux configuration from $U(1)_0$ to $\pi-$flux, $\pi\pi 0 0$ flux or FF, as shown in Fig.~\ref{fig:magnetic_field_tuning}. The photon dynamics for the $\pi\pi00$ and $\pi$-flux phases, depicted in Fig.~\ref{fig:analytics+numerics}, are equivalent to the standard 0-flux $U(1)_0$ spin liquid (see SM~\cite{SM} for details).
However, this is not the case for the FF phase, which does not exhibit quantized flux, as can be seen in Fig.~\ref{fig:magnetic_field_tuning} -- this is a crucial distinction from the other three phases, as it reflects frustration of the emergent magnetic fluxes that gives rise to new behaviors. 

At higher temperatures, the nonlinearities of the system are exposed by the stronger fluctuations. Working in terms of the emergent large-$S$ magnetic flux $\Phi$, where the hexagon-flip operator $\mathcal{O} \sim e^{i\Phi}$, we see two contributions from the order of $\Phi^4$ terms: (i) a Hartree-Fock correction to the dispersion, and (ii) a lifetime broadening of the photon modes.
More subtly, we see a high temperature effect in the FF phase (Fig.~\ref{fig:analytics+numerics}d) that is absent in the non-frustrated phases (panel c): the appearance of a structured continuum of excitations above the photon bandwidth. In the language of nonlinear optics, this behaviour is referred to as the three-photon up-conversion (see SM~\cite{SM} for details).

\prlsection{Spinon dispersion}
\as{
The spinon spectra of different QSI phases are also of great importance, since they are observable by inelastic neutron scattering experiments~\cite{gao2024emergent,poree2023fractional}. 
Subtly, $\boldsymbol{B}$ exerts negligible force on the octupolar spinons -- instead, it modulates their nearest-neighbour hopping amplitude $\hat{\vb*{z}}_i\cdot \vb*{B}$, thereby splitting the degenerate spinon levels by an energy scale $\propto \|\vb*{B}\|$. 
}

\as{Here we present the spinon spectra in the frustrated-flux and $\pi\pi00$ phases, and highlight their very different phenomenology from the $0$-flux and $\pi$-flux phase.  
The calculation is done using gauge mean field theory (GMFT)
~\cite{chenSpectralPeriodicitySpinon2017,chenDistinguishingThermodynamicsSpectroscopy2023,yaoPyrochloreSpinLiquid2020,chenDiracMagneticMonopole2017,liSymmetryEnrichedTopological2017},
\as{which has been shown to produce results in good agreements with experiments and numerical simulations~\cite{desrochersSpectroscopicSignaturesFractionalization2023,desrochersSymmetryFractionalizationGauge2023}.}
}

\begin{figure}
    \includegraphics[width=\columnwidth]{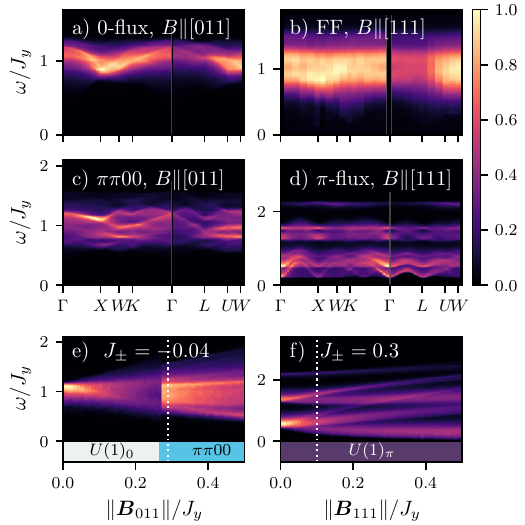}
    \caption{\as{GMFT spinon spectral weight for DO-QSI under applied magnetic field. Panels a)-d) show momentum-resolved structure factors, panels e) and f) show Brillouin zone integrated intensity as a function of external field strength. Parameters used: a) $J_\pm = -0.04, \|\boldsymbol{B}\|=0.06$, b) $J_\pm =-0.04, \|\boldsymbol{B}\|=0.29$, c) $J_\pm = -0.04, \|\boldsymbol{B}\|=0.29$, d) $J_\pm = 0.30, \|\boldsymbol{B}\|=0.10$. Vertical dotted lines in panels e) and f) correspond to the parameter sets of panels c) and d), respectively. Note differing $\omega$-axis scales.
    }}
    \label{fig:spinon-main}
\end{figure}

\as{The two-spinon spectra, shown in Fig.~\ref{fig:spinon-main}, distinguish the four phases in two ways: i) the momentum-resolved distribution of spectral weight is noticeably different; 
and ii) the dependence of this spectral weight on the applied field varies greatly between phases.
We find the splitting of the spinon branches in the $\pi$-flux phase to be largely independent of the applied magnetic field alignment with the crystal~\cite{SM}, suggesting that measurements on powders could yield results similar to Fig.~\ref{fig:spinon-main}f.
Our result for the $\pi\pi00$ phase is consistent with Ref.~\onlinecite{zhou2024arxiv}.
}

\as{
The FF phase is distinguished from all other phases by a very small Brillouin zone, and consequently a large number of nearly-flat bands (see SM~\cite{SM} and Fig.~\ref{fig:spinon-main}b). 
General magnetic fields may in principle give rise to a fully incommensurate gauge structure. We therefore expect the spinon spectrum to resemble a 3D Hofstadter butterfly~\cite{parkGeometricSuperconductivity3D2020,hofstadterEnergyLevelsWave1976}. 
In addition, spinon bands in the FF phase are known to have finite Chern number, yielding experimental signatures for instance in the thermal Hall effect~\cite{yangMagneticFieldThermal2020}.}

\prlsection{Magnetic field tuning in DO-QSI vs QSI}
We emphasize  the differences between DO-QSI and QSI in regards to the rich physics discussed above. In conventional QSI, applied magnetic fields unavoidably couple to the component of the spins subjected to the ice rules~\footnote{Note that, due to the 3D non-collinear nature of the local easy axes, it is not possible to apply a purely transverse external field.} and favour a polarised, classical ice-state as a result. This selection is only countered by the kinetic terms in the Hamiltonian, of the order of $\Jring\vert_{\vb*{B}=0} \sim J_\pm^3 / J_z^2$ and much smaller than the dominant exchange coupling. One would therefore be limited in the use of applied fields to tune the eQED behaviour to very small field strengths. 
While some precursors of the behaviours discussed in our work may be achievable in conventional QSI, it is likely that they would remain very subtle; one would not be able to observe the rich phase diagram and, for instance, the exotic physics of the FF phase at large fields ($\vert \vb*{B} \vert \gtrsim \sqrt{|J_\pm| J_z} \gg \Jring\vert_{\vb*{B}=0}$). 

In contrast, DO-QSI is free from such a problem, since the field does not couple to the dominant octupolar component $\sigma^z$ subjected to the ice-rule, and acts as a purely transverse field. Hence, a phase transition out of the eQED phase is only expected for $\vert \vb*{B} \vert \gtrsim J_y$~\cite{liSymmetryEnrichedTopological2017}. 
This is crucial for the ability to access the predicted tunable eQED regime in experiments. 

Among the DO-QSI candidates uncovered to date (summarized for convenience in Table~\ref{tab:materials} in the SM~\cite{SM}), it has been suggested that the compounds $\mathrm{Ce}_2\mathrm{Hf}_2\mathrm{O}_7$ and $\mathrm{Ce}_2\mathrm{Sn}_2\mathrm{O}_7$ may have dominant $J_y$ interactions, while $\mathrm{Ce}_2\mathrm{Zr}_2\mathrm{O}_7$ has comparable $J_x$ and $J_y$. 
Our proposal is thus directly applicable to the former two materials, for which single crystals exist.

\prlsection{Discussion}
In this work, we demonstrated that various interesting eQED physics in DO-QSI systems can be accurately tuned using experimentally accessible applied magnetic fields. Modest field strengths ($\lesssim \sqrt{|J_\pm|J_y}$) generically cause sizable, anisotropic changes to both photon and spinon dispersions.
Strong fields in the $U(1)_0$ phase of DO-QSI have the even more dramatic effect of driving the system into eQED phases with different background flux configurations. The frustrated-flux phase is particularly unusual, as it allows for the presence of nonlinear three-photon processes that qualitatively alter the photon dispersion. 
By-design emergent-photon dynamics also opens the gate to other possible interesting photonic phenomena including 
lensing effects and wave guides based on total internal reflection.

\as{Our GMFT spinon calculations suggest that the four phases may be straightforwardly identified through their fingerprints in spin correlations, and in particular comparing how they evolve with magnetic field. These signatures remain fairly clear even when integrated across the Brillouin zone, or indeed restricted to the $\Gamma$ point, rendering them accessible to a wide variety of experimental techniques.} 
Our proposal can be straightforwardly applied to DO-QSI candidate materials with dominant $J_y$ interactions.

In addition, the Zeeman coupling $\vb*{\hat{z}}_i\cdot\vb*{B} \sigma^x_i$ contributes to spinon propagation with the subtle difference that it hops spinons between nearest-neighbour tetrahedra, instead of the next nearest neighbour ones (as due to the $J_\pm \sigma^+\sigma^-$ terms). 
The latter preserve the spinon number on the  `up' and `down' tetrahedral sublattice separately, while the former couples the two sublattices and lifts the spinon band degeneracy~\cite{liSymmetryEnrichedTopological2017}.
This suggests that by applying a magnetic field, one can introduce net spinon charge imbalances between the two tetrahedral sublattices. Later quenching the field to zero, one can abruptly decouple the populations of `up' and `down' spinons, locking in the charge imbalance. 
Such a procedure may be used to generate QSI states hosting spinons and antispinons (on different tetrahedral sublattices) that are unable to annihilate and therefore survive down to vanishing temperatures, given proper boundary conditions. 

Another interesting perspective comes from the eQED field theory discussed in Refs.~\onlinecite{paceEmergentFineStructure2021,laumannHybridDyonsInverted2023,morampudi2020}. With the photon speed controlled by the applied magnetic field, we expect the fine-structure constant to be tunable too. This provides an exciting avenue to probe both weakly and strongly coupled eQED, ultrarelativistic effects, and transitions between different regimes. 
Emergent light by-design may soon become a reality in DO-QSI, presenting us with a new landscape of QED beyond elementary particle physics to be explored. 

\begin{acknowledgments}
The authors would like to thank
Gang Chen and Yamashita Minoru for helpful discussions. 
This work was funded in part by the Engineering and Physical Sciences Research Council (EPSRC) grants No.~EP/T028580/1 and No.~EP/V062654/1 (CC). AHN and HY are grateful for the hospitality of the Aspen Center for Physics, supported by the U.S. National Science Foundation grant PHY-2210452. AS would like to gratefully acknowledge funding from Cambridge Australia Scholarships.
HY is supported by 2024 Toyota Riken Scholar Program  from Toyota Physical and Chemical Research Institute.
\end{acknowledgments}

\bibliography{ref.bib}{}

\begin{thebibliography}{49}%
\makeatletter
\providecommand \@ifxundefined [1]{%
 \@ifx{#1\undefined}
}%
\providecommand \@ifnum [1]{%
 \ifnum #1\expandafter \@firstoftwo
 \else \expandafter \@secondoftwo
 \fi
}%
\providecommand \@ifx [1]{%
 \ifx #1\expandafter \@firstoftwo
 \else \expandafter \@secondoftwo
 \fi
}%
\providecommand \natexlab [1]{#1}%
\providecommand \enquote  [1]{``#1''}%
\providecommand \bibnamefont  [1]{#1}%
\providecommand \bibfnamefont [1]{#1}%
\providecommand \citenamefont [1]{#1}%
\providecommand \href@noop [0]{\@secondoftwo}%
\providecommand \href [0]{\begingroup \@sanitize@url \@href}%
\providecommand \@href[1]{\@@startlink{#1}\@@href}%
\providecommand \@@href[1]{\endgroup#1\@@endlink}%
\providecommand \@sanitize@url [0]{\catcode `\\12\catcode `\$12\catcode `\&12\catcode `\#12\catcode `\^12\catcode `\_12\catcode `\%12\relax}%
\providecommand \@@startlink[1]{}%
\providecommand \@@endlink[0]{}%
\providecommand \url  [0]{\begingroup\@sanitize@url \@url }%
\providecommand \@url [1]{\endgroup\@href {#1}{\urlprefix }}%
\providecommand \urlprefix  [0]{URL }%
\providecommand \Eprint [0]{\href }%
\providecommand \doibase [0]{https://doi.org/}%
\providecommand \selectlanguage [0]{\@gobble}%
\providecommand \bibinfo  [0]{\@secondoftwo}%
\providecommand \bibfield  [0]{\@secondoftwo}%
\providecommand \translation [1]{[#1]}%
\providecommand \BibitemOpen [0]{}%
\providecommand \bibitemStop [0]{}%
\providecommand \bibitemNoStop [0]{.\EOS\space}%
\providecommand \EOS [0]{\spacefactor3000\relax}%
\providecommand \BibitemShut  [1]{\csname bibitem#1\endcsname}%
\let\auto@bib@innerbib\@empty
\bibitem [{\citenamefont {Moessner}\ and\ \citenamefont {Moore}(2021)}]{moessnerTopologicalPhasesMatter2021}%
  \BibitemOpen
  \bibfield  {author} {\bibinfo {author} {\bibfnamefont {R.}~\bibnamefont {Moessner}}\ and\ \bibinfo {author} {\bibfnamefont {J.~E.}\ \bibnamefont {Moore}},\ }\href {https://doi.org/10.1017/9781316226308} {\emph {\bibinfo {title} {Topological {{Phases}} of {{Matter}}}}}\ (\bibinfo  {publisher} {{Cambridge University Press}},\ \bibinfo {year} {2021})\BibitemShut {NoStop}%
\bibitem [{\citenamefont {Gingras}\ and\ \citenamefont {McClarty}(2014)}]{gingrasQuantumSpinIce2014}%
  \BibitemOpen
  \bibfield  {author} {\bibinfo {author} {\bibfnamefont {M.~J.~P.}\ \bibnamefont {Gingras}}\ and\ \bibinfo {author} {\bibfnamefont {P.~A.}\ \bibnamefont {McClarty}},\ }\href {https://doi.org/10.1088/0034-4885/77/5/056501} {\bibfield  {journal} {\bibinfo  {journal} {Reports on Progress in Physics}\ }\textbf {\bibinfo {volume} {77}},\ \bibinfo {pages} {056501} (\bibinfo {year} {2014})}\BibitemShut {NoStop}%
\bibitem [{\citenamefont {Udagawa}\ and\ \citenamefont {Jaubert}(2021)}]{udagawaSpinIce2021}%
  \BibitemOpen
  \bibinfo {editor} {\bibfnamefont {M.}~\bibnamefont {Udagawa}}\ and\ \bibinfo {editor} {\bibfnamefont {L.}~\bibnamefont {Jaubert}},\ eds.,\ \href {https://doi.org/10.1007/978-3-030-70860-3} {\emph {\bibinfo {title} {Spin {{Ice}}}}},\ \bibinfo {series} {Springer {{Series}} in {{Solid-State Sciences}}}, Vol.\ \bibinfo {volume} {197}\ (\bibinfo  {publisher} {{Springer International Publishing}},\ \bibinfo {year} {2021})\BibitemShut {NoStop}%
\bibitem [{\citenamefont {Hermele}\ \emph {et~al.}(2004)\citenamefont {Hermele}, \citenamefont {Fisher},\ and\ \citenamefont {Balents}}]{Hermele_Fisher_Balents}%
  \BibitemOpen
  \bibfield  {author} {\bibinfo {author} {\bibfnamefont {M.}~\bibnamefont {Hermele}}, \bibinfo {author} {\bibfnamefont {M.~P.~A.}\ \bibnamefont {Fisher}},\ and\ \bibinfo {author} {\bibfnamefont {L.}~\bibnamefont {Balents}},\ }\href {https://doi.org/10.1103/PhysRevB.69.064404} {\bibfield  {journal} {\bibinfo  {journal} {Phys. Rev. B}\ }\textbf {\bibinfo {volume} {69}},\ \bibinfo {pages} {064404} (\bibinfo {year} {2004})}\BibitemShut {NoStop}%
\bibitem [{\citenamefont {Huang}\ \emph {et~al.}(2018)\citenamefont {Huang}, \citenamefont {Deng}, \citenamefont {Wan},\ and\ \citenamefont {Meng}}]{huangDynamicsTopologicalExcitations2018}%
  \BibitemOpen
  \bibfield  {author} {\bibinfo {author} {\bibfnamefont {C.-J.}\ \bibnamefont {Huang}}, \bibinfo {author} {\bibfnamefont {Y.}~\bibnamefont {Deng}}, \bibinfo {author} {\bibfnamefont {Y.}~\bibnamefont {Wan}},\ and\ \bibinfo {author} {\bibfnamefont {Z.~Y.}\ \bibnamefont {Meng}},\ }\href {https://doi.org/10.1103/PhysRevLett.120.167202} {\bibfield  {journal} {\bibinfo  {journal} {Physical Review Letters}\ }\textbf {\bibinfo {volume} {120}},\ \bibinfo {pages} {167202} (\bibinfo {year} {2018})}\BibitemShut {NoStop}%
\bibitem [{\citenamefont {Shannon}\ \emph {et~al.}(2012)\citenamefont {Shannon}, \citenamefont {Sikora}, \citenamefont {Pollmann}, \citenamefont {Penc},\ and\ \citenamefont {Fulde}}]{shannonQuantumIceQuantum2012}%
  \BibitemOpen
  \bibfield  {author} {\bibinfo {author} {\bibfnamefont {N.}~\bibnamefont {Shannon}}, \bibinfo {author} {\bibfnamefont {O.}~\bibnamefont {Sikora}}, \bibinfo {author} {\bibfnamefont {F.}~\bibnamefont {Pollmann}}, \bibinfo {author} {\bibfnamefont {K.}~\bibnamefont {Penc}},\ and\ \bibinfo {author} {\bibfnamefont {P.}~\bibnamefont {Fulde}},\ }\href {https://doi.org/10.1103/PhysRevLett.108.067204} {\bibfield  {journal} {\bibinfo  {journal} {Physical Review Letters}\ }\textbf {\bibinfo {volume} {108}},\ \bibinfo {pages} {067204} (\bibinfo {year} {2012})}\BibitemShut {NoStop}%
\bibitem [{\citenamefont {Udagawa}\ and\ \citenamefont {Moessner}(2019)}]{udagawaSpectrumItinerantFractional2019}%
  \BibitemOpen
  \bibfield  {author} {\bibinfo {author} {\bibfnamefont {M.}~\bibnamefont {Udagawa}}\ and\ \bibinfo {author} {\bibfnamefont {R.}~\bibnamefont {Moessner}},\ }\href {https://doi.org/10.1103/PhysRevLett.122.117201} {\bibfield  {journal} {\bibinfo  {journal} {Physical Review Letters}\ }\textbf {\bibinfo {volume} {122}},\ \bibinfo {pages} {117201} (\bibinfo {year} {2019})}\BibitemShut {NoStop}%
\bibitem [{\citenamefont {Hosoi}\ \emph {et~al.}(2022)\citenamefont {Hosoi}, \citenamefont {Zhang}, \citenamefont {Patri},\ and\ \citenamefont {Kim}}]{hosoiUncoveringFootprintsDipolarOctupolar2022}%
  \BibitemOpen
  \bibfield  {author} {\bibinfo {author} {\bibfnamefont {M.}~\bibnamefont {Hosoi}}, \bibinfo {author} {\bibfnamefont {E.~Z.}\ \bibnamefont {Zhang}}, \bibinfo {author} {\bibfnamefont {A.~S.}\ \bibnamefont {Patri}},\ and\ \bibinfo {author} {\bibfnamefont {Y.~B.}\ \bibnamefont {Kim}},\ }\href {https://doi.org/10.1103/PhysRevLett.129.097202} {\bibfield  {journal} {\bibinfo  {journal} {Physical Review Letters}\ }\textbf {\bibinfo {volume} {129}},\ \bibinfo {pages} {097202} (\bibinfo {year} {2022})}\BibitemShut {NoStop}%
\bibitem [{\citenamefont {Benton}\ \emph {et~al.}(2012)\citenamefont {Benton}, \citenamefont {Sikora},\ and\ \citenamefont {Shannon}}]{BentonPhysRevB}%
  \BibitemOpen
  \bibfield  {author} {\bibinfo {author} {\bibfnamefont {O.}~\bibnamefont {Benton}}, \bibinfo {author} {\bibfnamefont {O.}~\bibnamefont {Sikora}},\ and\ \bibinfo {author} {\bibfnamefont {N.}~\bibnamefont {Shannon}},\ }\href {https://doi.org/10.1103/PhysRevB.86.075154} {\bibfield  {journal} {\bibinfo  {journal} {Phys. Rev. B}\ }\textbf {\bibinfo {volume} {86}},\ \bibinfo {pages} {075154} (\bibinfo {year} {2012})}\BibitemShut {NoStop}%
\bibitem [{\citenamefont {Pace}\ \emph {et~al.}(2021)\citenamefont {Pace}, \citenamefont {Morampudi}, \citenamefont {Moessner},\ and\ \citenamefont {Laumann}}]{paceEmergentFineStructure2021}%
  \BibitemOpen
  \bibfield  {author} {\bibinfo {author} {\bibfnamefont {S.~D.}\ \bibnamefont {Pace}}, \bibinfo {author} {\bibfnamefont {S.~C.}\ \bibnamefont {Morampudi}}, \bibinfo {author} {\bibfnamefont {R.}~\bibnamefont {Moessner}},\ and\ \bibinfo {author} {\bibfnamefont {C.~R.}\ \bibnamefont {Laumann}},\ }\href {https://doi.org/10.1103/PhysRevLett.127.117205} {\bibfield  {journal} {\bibinfo  {journal} {Physical Review Letters}\ }\textbf {\bibinfo {volume} {127}},\ \bibinfo {pages} {117205} (\bibinfo {year} {2021})}\BibitemShut {NoStop}%
\bibitem [{\citenamefont {Chen}(2023)}]{chenDistinguishingThermodynamicsSpectroscopy2023}%
  \BibitemOpen
  \bibfield  {author} {\bibinfo {author} {\bibfnamefont {G.}~\bibnamefont {Chen}},\ }\href {https://doi.org/10.1103/PhysRevResearch.5.033169} {\bibfield  {journal} {\bibinfo  {journal} {Physical Review Research}\ }\textbf {\bibinfo {volume} {5}},\ \bibinfo {pages} {033169} (\bibinfo {year} {2023})}\BibitemShut {NoStop}%
\bibitem [{\citenamefont {Desrochers}\ and\ \citenamefont {Kim}(2024)}]{desrochersSpectroscopicSignaturesFractionalization2023}%
  \BibitemOpen
  \bibfield  {author} {\bibinfo {author} {\bibfnamefont {F.}~\bibnamefont {Desrochers}}\ and\ \bibinfo {author} {\bibfnamefont {Y.~B.}\ \bibnamefont {Kim}},\ }\href {https://journals.aps.org/prl/abstract/10.1103/PhysRevLett.132.066502} {\bibfield  {journal} {\bibinfo  {journal} {Phys. Rev. Lett.}\ }\textbf {\bibinfo {volume} {132}},\ \bibinfo {pages} {066502} (\bibinfo {year} {2024})}\BibitemShut {NoStop}%
\bibitem [{\citenamefont {Lee}\ \emph {et~al.}(2012)\citenamefont {Lee}, \citenamefont {Onoda},\ and\ \citenamefont {Balents}}]{leeGenericQuantumSpin2012}%
  \BibitemOpen
  \bibfield  {author} {\bibinfo {author} {\bibfnamefont {S.}~\bibnamefont {Lee}}, \bibinfo {author} {\bibfnamefont {S.}~\bibnamefont {Onoda}},\ and\ \bibinfo {author} {\bibfnamefont {L.}~\bibnamefont {Balents}},\ }\href {https://doi.org/10.1103/PhysRevB.86.104412} {\bibfield  {journal} {\bibinfo  {journal} {Physical Review B}\ }\textbf {\bibinfo {volume} {86}},\ \bibinfo {pages} {104412} (\bibinfo {year} {2012})}\BibitemShut {NoStop}%
\bibitem [{\citenamefont {Savary}\ and\ \citenamefont {Balents}(2012)}]{savaryCoulombicQuantumLiquids2012}%
  \BibitemOpen
  \bibfield  {author} {\bibinfo {author} {\bibfnamefont {L.}~\bibnamefont {Savary}}\ and\ \bibinfo {author} {\bibfnamefont {L.}~\bibnamefont {Balents}},\ }\href {https://doi.org/10.1103/PhysRevLett.108.037202} {\bibfield  {journal} {\bibinfo  {journal} {Physical Review Letters}\ }\textbf {\bibinfo {volume} {108}},\ \bibinfo {pages} {037202} (\bibinfo {year} {2012})}\BibitemShut {NoStop}%
\bibitem [{\citenamefont {Morampudi}\ \emph {et~al.}(2020)\citenamefont {Morampudi}, \citenamefont {Wilczek},\ and\ \citenamefont {Laumann}}]{morampudi2020}%
  \BibitemOpen
  \bibfield  {author} {\bibinfo {author} {\bibfnamefont {S.~C.}\ \bibnamefont {Morampudi}}, \bibinfo {author} {\bibfnamefont {F.}~\bibnamefont {Wilczek}},\ and\ \bibinfo {author} {\bibfnamefont {C.~R.}\ \bibnamefont {Laumann}},\ }\href {https://doi.org/10.1103/PhysRevLett.124.097204} {\bibfield  {journal} {\bibinfo  {journal} {Physical Review Letters}\ }\textbf {\bibinfo {volume} {124}},\ \bibinfo {pages} {097204} (\bibinfo {year} {2020})}\BibitemShut {NoStop}%
\bibitem [{\citenamefont {Szab\'o}\ and\ \citenamefont {Castelnovo}(2019)}]{szaboSeeingLightVison2019a}%
  \BibitemOpen
  \bibfield  {author} {\bibinfo {author} {\bibfnamefont {A.}~\bibnamefont {Szab\'o}}\ and\ \bibinfo {author} {\bibfnamefont {C.}~\bibnamefont {Castelnovo}},\ }\href {https://doi.org/10.1103/PhysRevB.100.014417} {\bibfield  {journal} {\bibinfo  {journal} {Physical Review B}\ }\textbf {\bibinfo {volume} {100}},\ \bibinfo {pages} {014417} (\bibinfo {year} {2019})}\BibitemShut {NoStop}%
\bibitem [{\citenamefont {Kwasigroch}(2020)}]{kwasigrochVisongeneratedPhotonMass2020}%
  \BibitemOpen
  \bibfield  {author} {\bibinfo {author} {\bibfnamefont {M.~P.}\ \bibnamefont {Kwasigroch}},\ }\href {https://doi.org/10.1103/PhysRevB.102.125113} {\bibfield  {journal} {\bibinfo  {journal} {Physical Review B}\ }\textbf {\bibinfo {volume} {102}},\ \bibinfo {pages} {125113} (\bibinfo {year} {2020})}\BibitemShut {NoStop}%
\bibitem [{\citenamefont {Sanders}\ and\ \citenamefont {Castelnovo}(2024)}]{sandersVisonCrystalQuantum2023}%
  \BibitemOpen
  \bibfield  {author} {\bibinfo {author} {\bibfnamefont {A.}~\bibnamefont {Sanders}}\ and\ \bibinfo {author} {\bibfnamefont {C.}~\bibnamefont {Castelnovo}},\ }\href {https://doi.org/10.1103/PhysRevB.109.094426} {\bibfield  {journal} {\bibinfo  {journal} {Phys. Rev. B}\ }\textbf {\bibinfo {volume} {109}},\ \bibinfo {pages} {094426} (\bibinfo {year} {2024})}\BibitemShut {NoStop}%
\bibitem [{\citenamefont {Chen}(2017{\natexlab{a}})}]{chenDiracMagneticMonopole2017}%
  \BibitemOpen
  \bibfield  {author} {\bibinfo {author} {\bibfnamefont {G.}~\bibnamefont {Chen}},\ }\href {https://doi.org/10.1103/PhysRevB.96.195127} {\bibfield  {journal} {\bibinfo  {journal} {Physical Review B}\ }\textbf {\bibinfo {volume} {96}},\ \bibinfo {pages} {195127} (\bibinfo {year} {2017}{\natexlab{a}})}\BibitemShut {NoStop}%
\bibitem [{\citenamefont {Bergman}\ \emph {et~al.}(2006)\citenamefont {Bergman}, \citenamefont {Fiete},\ and\ \citenamefont {Balents}}]{bergmanOrderingFrustratedPyrochlore2006}%
  \BibitemOpen
  \bibfield  {author} {\bibinfo {author} {\bibfnamefont {D.~L.}\ \bibnamefont {Bergman}}, \bibinfo {author} {\bibfnamefont {G.~A.}\ \bibnamefont {Fiete}},\ and\ \bibinfo {author} {\bibfnamefont {L.}~\bibnamefont {Balents}},\ }\href {https://doi.org/10.1103/PhysRevB.73.134402} {\bibfield  {journal} {\bibinfo  {journal} {Physical Review B}\ }\textbf {\bibinfo {volume} {73}},\ \bibinfo {pages} {134402} (\bibinfo {year} {2006})}\BibitemShut {NoStop}%
\bibitem [{\citenamefont {Huang}\ \emph {et~al.}(2014)\citenamefont {Huang}, \citenamefont {Chen},\ and\ \citenamefont {Hermele}}]{HuangPhysRevLett}%
  \BibitemOpen
  \bibfield  {author} {\bibinfo {author} {\bibfnamefont {Y.-P.}\ \bibnamefont {Huang}}, \bibinfo {author} {\bibfnamefont {G.}~\bibnamefont {Chen}},\ and\ \bibinfo {author} {\bibfnamefont {M.}~\bibnamefont {Hermele}},\ }\href {https://doi.org/10.1103/PhysRevLett.112.167203} {\bibfield  {journal} {\bibinfo  {journal} {Phys. Rev. Lett.}\ }\textbf {\bibinfo {volume} {112}},\ \bibinfo {pages} {167203} (\bibinfo {year} {2014})}\BibitemShut {NoStop}%
\bibitem [{\citenamefont {Gao}\ \emph {et~al.}(2024)\citenamefont {Gao}, \citenamefont {Desrochers}, \citenamefont {Tam}, \citenamefont {Steffens}, \citenamefont {Hiess}, \citenamefont {Su}, \citenamefont {Cheong}, \citenamefont {Kim},\ and\ \citenamefont {Dai}}]{gao2024emergent}%
  \BibitemOpen
  \bibfield  {author} {\bibinfo {author} {\bibfnamefont {B.}~\bibnamefont {Gao}}, \bibinfo {author} {\bibfnamefont {F.}~\bibnamefont {Desrochers}}, \bibinfo {author} {\bibfnamefont {D.~W.}\ \bibnamefont {Tam}}, \bibinfo {author} {\bibfnamefont {P.}~\bibnamefont {Steffens}}, \bibinfo {author} {\bibfnamefont {A.}~\bibnamefont {Hiess}}, \bibinfo {author} {\bibfnamefont {Y.}~\bibnamefont {Su}}, \bibinfo {author} {\bibfnamefont {S.-W.}\ \bibnamefont {Cheong}}, \bibinfo {author} {\bibfnamefont {Y.~B.}\ \bibnamefont {Kim}},\ and\ \bibinfo {author} {\bibfnamefont {P.}~\bibnamefont {Dai}},\ }\href@noop {} {\bibinfo {title} {Emergent photons and fractionalized excitations in a quantum spin liquid}} (\bibinfo {year} {2024}),\ \Eprint {https://arxiv.org/abs/2404.04207} {arXiv:2404.04207 [cond-mat.str-el]} \BibitemShut {NoStop}%
\bibitem [{\citenamefont {Gao}\ \emph {et~al.}(2019)\citenamefont {Gao}, \citenamefont {Chen}, \citenamefont {Tam}, \citenamefont {Huang}, \citenamefont {Sasmal}, \citenamefont {Adroja}, \citenamefont {Ye}, \citenamefont {Cao}, \citenamefont {Sala}, \citenamefont {Stone}, \citenamefont {Baines}, \citenamefont {Verezhak}, \citenamefont {Hu}, \citenamefont {Chung}, \citenamefont {Xu}, \citenamefont {Cheong}, \citenamefont {Nallaiyan}, \citenamefont {Spagna}, \citenamefont {Maple}, \citenamefont {Nevidomskyy}, \citenamefont {Morosan}, \citenamefont {Chen},\ and\ \citenamefont {Dai}}]{CZO.Gao}%
  \BibitemOpen
  \bibfield  {author} {\bibinfo {author} {\bibfnamefont {B.}~\bibnamefont {Gao}}, \bibinfo {author} {\bibfnamefont {T.}~\bibnamefont {Chen}}, \bibinfo {author} {\bibfnamefont {D.~W.}\ \bibnamefont {Tam}}, \bibinfo {author} {\bibfnamefont {C.-L.}\ \bibnamefont {Huang}}, \bibinfo {author} {\bibfnamefont {K.}~\bibnamefont {Sasmal}}, \bibinfo {author} {\bibfnamefont {D.~T.}\ \bibnamefont {Adroja}}, \bibinfo {author} {\bibfnamefont {F.}~\bibnamefont {Ye}}, \bibinfo {author} {\bibfnamefont {H.}~\bibnamefont {Cao}}, \bibinfo {author} {\bibfnamefont {G.}~\bibnamefont {Sala}}, \bibinfo {author} {\bibfnamefont {M.~B.}\ \bibnamefont {Stone}}, \bibinfo {author} {\bibfnamefont {C.}~\bibnamefont {Baines}}, \bibinfo {author} {\bibfnamefont {J.~A.~T.}\ \bibnamefont {Verezhak}}, \bibinfo {author} {\bibfnamefont {H.}~\bibnamefont {Hu}}, \bibinfo {author} {\bibfnamefont {J.-H.}\ \bibnamefont {Chung}}, \bibinfo {author} {\bibfnamefont {X.}~\bibnamefont {Xu}}, \bibinfo {author} {\bibfnamefont {S.-W.}\ \bibnamefont {Cheong}}, \bibinfo {author} {\bibfnamefont {M.}~\bibnamefont {Nallaiyan}}, \bibinfo {author} {\bibfnamefont {S.}~\bibnamefont {Spagna}}, \bibinfo {author} {\bibfnamefont {M.~B.}\ \bibnamefont {Maple}}, \bibinfo {author} {\bibfnamefont {A.~H.}\ \bibnamefont {Nevidomskyy}}, \bibinfo {author} {\bibfnamefont {E.}~\bibnamefont {Morosan}}, \bibinfo {author} {\bibfnamefont {G.}~\bibnamefont {Chen}},\ and\ \bibinfo {author} {\bibfnamefont {P.}~\bibnamefont {Dai}},\ }\href {https://doi.org/10.1038/s41567-019-0577-6} {\bibfield  {journal} {\bibinfo  {journal} {Nature Physics}\ }\textbf {\bibinfo {volume} {15}},\ \bibinfo {pages} {1052} (\bibinfo {year} {2019})}\BibitemShut {NoStop}%
\bibitem [{\citenamefont {Gaudet}\ \emph {et~al.}(2019)\citenamefont {Gaudet}, \citenamefont {Smith}, \citenamefont {Dudemaine}, \citenamefont {Beare}, \citenamefont {Buhariwalla}, \citenamefont {Butch}, \citenamefont {Stone}, \citenamefont {Kolesnikov}, \citenamefont {Xu}, \citenamefont {Yahne}, \citenamefont {Ross}, \citenamefont {Marjerrison}, \citenamefont {Garrett}, \citenamefont {Luke}, \citenamefont {Bianchi},\ and\ \citenamefont {Gaulin}}]{CZO.Gaulin}%
  \BibitemOpen
  \bibfield  {author} {\bibinfo {author} {\bibfnamefont {J.}~\bibnamefont {Gaudet}}, \bibinfo {author} {\bibfnamefont {E.~M.}\ \bibnamefont {Smith}}, \bibinfo {author} {\bibfnamefont {J.}~\bibnamefont {Dudemaine}}, \bibinfo {author} {\bibfnamefont {J.}~\bibnamefont {Beare}}, \bibinfo {author} {\bibfnamefont {C.~R.~C.}\ \bibnamefont {Buhariwalla}}, \bibinfo {author} {\bibfnamefont {N.~P.}\ \bibnamefont {Butch}}, \bibinfo {author} {\bibfnamefont {M.~B.}\ \bibnamefont {Stone}}, \bibinfo {author} {\bibfnamefont {A.~I.}\ \bibnamefont {Kolesnikov}}, \bibinfo {author} {\bibfnamefont {G.}~\bibnamefont {Xu}}, \bibinfo {author} {\bibfnamefont {D.~R.}\ \bibnamefont {Yahne}}, \bibinfo {author} {\bibfnamefont {K.~A.}\ \bibnamefont {Ross}}, \bibinfo {author} {\bibfnamefont {C.~A.}\ \bibnamefont {Marjerrison}}, \bibinfo {author} {\bibfnamefont {J.~D.}\ \bibnamefont {Garrett}}, \bibinfo {author} {\bibfnamefont {G.~M.}\ \bibnamefont {Luke}}, \bibinfo {author} {\bibfnamefont {A.~D.}\ \bibnamefont {Bianchi}},\ and\ \bibinfo {author} {\bibfnamefont {B.~D.}\ \bibnamefont {Gaulin}},\ }\href {https://doi.org/10.1103/PhysRevLett.122.187201} {\bibfield  {journal} {\bibinfo  {journal} {Phys. Rev. Lett.}\ }\textbf {\bibinfo {volume} {122}},\ \bibinfo {pages} {187201} (\bibinfo {year} {2019})}\BibitemShut {NoStop}%
\bibitem [{\citenamefont {Por\'ee}\ \emph {et~al.}(2022)\citenamefont {Por\'ee}, \citenamefont {Lhotel}, \citenamefont {Petit}, \citenamefont {Krajewska}, \citenamefont {Puphal}, \citenamefont {Clark}, \citenamefont {Pomjakushin}, \citenamefont {Walker}, \citenamefont {Gauthier}, \citenamefont {Gawryluk},\ and\ \citenamefont {Sibille}}]{poreeCrystalfieldStatesDefect2022}%
  \BibitemOpen
  \bibfield  {author} {\bibinfo {author} {\bibfnamefont {V.}~\bibnamefont {Por\'ee}}, \bibinfo {author} {\bibfnamefont {E.}~\bibnamefont {Lhotel}}, \bibinfo {author} {\bibfnamefont {S.}~\bibnamefont {Petit}}, \bibinfo {author} {\bibfnamefont {A.}~\bibnamefont {Krajewska}}, \bibinfo {author} {\bibfnamefont {P.}~\bibnamefont {Puphal}}, \bibinfo {author} {\bibfnamefont {A.~H.}\ \bibnamefont {Clark}}, \bibinfo {author} {\bibfnamefont {V.}~\bibnamefont {Pomjakushin}}, \bibinfo {author} {\bibfnamefont {H.~C.}\ \bibnamefont {Walker}}, \bibinfo {author} {\bibfnamefont {N.}~\bibnamefont {Gauthier}}, \bibinfo {author} {\bibfnamefont {D.~J.}\ \bibnamefont {Gawryluk}},\ and\ \bibinfo {author} {\bibfnamefont {R.}~\bibnamefont {Sibille}},\ }\href {https://doi.org/10.1103/PhysRevMaterials.6.044406} {\bibfield  {journal} {\bibinfo  {journal} {Physical Review Materials}\ }\textbf {\bibinfo {volume} {6}},\ \bibinfo {pages} {044406} (\bibinfo {year} {2022})},\ \Eprint {https://arxiv.org/abs/2203.16295} {2203.16295} \BibitemShut {NoStop}%
\bibitem [{\citenamefont {Por{\'e}e}\ \emph {et~al.}(2023)\citenamefont {Por{\'e}e}, \citenamefont {Bhardwaj}, \citenamefont {Lhotel}, \citenamefont {Petit}, \citenamefont {Gauthier}, \citenamefont {Yan}, \citenamefont {Pomjakushin}, \citenamefont {Ollivier}, \citenamefont {Quilliam}, \citenamefont {Nevidomskyy}, \citenamefont {Changlani},\ and\ \citenamefont {Sibille}}]{poree2023dipolaroctupolar}%
  \BibitemOpen
  \bibfield  {author} {\bibinfo {author} {\bibfnamefont {V.}~\bibnamefont {Por{\'e}e}}, \bibinfo {author} {\bibfnamefont {A.}~\bibnamefont {Bhardwaj}}, \bibinfo {author} {\bibfnamefont {E.}~\bibnamefont {Lhotel}}, \bibinfo {author} {\bibfnamefont {S.}~\bibnamefont {Petit}}, \bibinfo {author} {\bibfnamefont {N.}~\bibnamefont {Gauthier}}, \bibinfo {author} {\bibfnamefont {H.}~\bibnamefont {Yan}}, \bibinfo {author} {\bibfnamefont {V.}~\bibnamefont {Pomjakushin}}, \bibinfo {author} {\bibfnamefont {J.}~\bibnamefont {Ollivier}}, \bibinfo {author} {\bibfnamefont {J.~A.}\ \bibnamefont {Quilliam}}, \bibinfo {author} {\bibfnamefont {A.~H.}\ \bibnamefont {Nevidomskyy}}, \bibinfo {author} {\bibfnamefont {H.~J.}\ \bibnamefont {Changlani}},\ and\ \bibinfo {author} {\bibfnamefont {R.}~\bibnamefont {Sibille}},\ }\href@noop {} {\  (\bibinfo {year} {2023})},\ \Eprint {https://arxiv.org/abs/2305.08261} {arXiv:2305.08261} \BibitemShut {NoStop}%
\bibitem [{\citenamefont {Bhardwaj}\ \emph {et~al.}(2024)\citenamefont {Bhardwaj}, \citenamefont {Por\'{e}e}, \citenamefont {Yan}, \citenamefont {Gauthier}, \citenamefont {Lhotel}, \citenamefont {Petit}, \citenamefont {Quilliam}, \citenamefont {Nevidomskyy}, \citenamefont {Sibille},\ and\ \citenamefont {Changlani}}]{bhardwaj2024arXiv}%
  \BibitemOpen
  \bibfield  {author} {\bibinfo {author} {\bibfnamefont {A.}~\bibnamefont {Bhardwaj}}, \bibinfo {author} {\bibfnamefont {V.}~\bibnamefont {Por\'{e}e}}, \bibinfo {author} {\bibfnamefont {H.}~\bibnamefont {Yan}}, \bibinfo {author} {\bibfnamefont {N.}~\bibnamefont {Gauthier}}, \bibinfo {author} {\bibfnamefont {E.}~\bibnamefont {Lhotel}}, \bibinfo {author} {\bibfnamefont {S.}~\bibnamefont {Petit}}, \bibinfo {author} {\bibfnamefont {J.~A.}\ \bibnamefont {Quilliam}}, \bibinfo {author} {\bibfnamefont {A.~H.}\ \bibnamefont {Nevidomskyy}}, \bibinfo {author} {\bibfnamefont {R.}~\bibnamefont {Sibille}},\ and\ \bibinfo {author} {\bibfnamefont {H.~J.}\ \bibnamefont {Changlani}},\ }\href {https://arxiv.org/abs/2402.08723} {\bibinfo {title} {Thermodynamics of the dipole-octupole pyrochlore magnet ce$_2$hf$_2$o$_{7}$ in applied magnetic fields}} (\bibinfo {year} {2024}),\ \Eprint {https://arxiv.org/abs/2402.08723} {arXiv:2402.08723 [cond-mat.str-el]} \BibitemShut {NoStop}%
\bibitem [{\citenamefont {Sibille}\ \emph {et~al.}(2020)\citenamefont {Sibille}, \citenamefont {Gauthier}, \citenamefont {Lhotel}, \citenamefont {Por\'ee}, \citenamefont {Pomjakushin}, \citenamefont {Ewings}, \citenamefont {Perring}, \citenamefont {Ollivier}, \citenamefont {Wildes}, \citenamefont {Ritter}, \citenamefont {Hansen}, \citenamefont {Keen}, \citenamefont {Nilsen}, \citenamefont {Keller}, \citenamefont {Petit},\ and\ \citenamefont {Fennell}}]{Sibille2020NatPhys}%
  \BibitemOpen
  \bibfield  {author} {\bibinfo {author} {\bibfnamefont {R.}~\bibnamefont {Sibille}}, \bibinfo {author} {\bibfnamefont {N.}~\bibnamefont {Gauthier}}, \bibinfo {author} {\bibfnamefont {E.}~\bibnamefont {Lhotel}}, \bibinfo {author} {\bibfnamefont {V.}~\bibnamefont {Por\'ee}}, \bibinfo {author} {\bibfnamefont {V.}~\bibnamefont {Pomjakushin}}, \bibinfo {author} {\bibfnamefont {R.~A.}\ \bibnamefont {Ewings}}, \bibinfo {author} {\bibfnamefont {T.~G.}\ \bibnamefont {Perring}}, \bibinfo {author} {\bibfnamefont {J.}~\bibnamefont {Ollivier}}, \bibinfo {author} {\bibfnamefont {A.}~\bibnamefont {Wildes}}, \bibinfo {author} {\bibfnamefont {C.}~\bibnamefont {Ritter}}, \bibinfo {author} {\bibfnamefont {T.~C.}\ \bibnamefont {Hansen}}, \bibinfo {author} {\bibfnamefont {D.~A.}\ \bibnamefont {Keen}}, \bibinfo {author} {\bibfnamefont {G.~J.}\ \bibnamefont {Nilsen}}, \bibinfo {author} {\bibfnamefont {L.}~\bibnamefont {Keller}}, \bibinfo {author} {\bibfnamefont {S.}~\bibnamefont {Petit}},\ and\ \bibinfo {author} {\bibfnamefont {T.}~\bibnamefont {Fennell}},\ }\href {https://doi.org/10.1038/s41567-020-0827-7} {\bibfield  {journal} {\bibinfo  {journal} {Nature Physics}\ }\textbf {\bibinfo {volume} {16}},\ \bibinfo {pages} {546} (\bibinfo {year} {2020})}\BibitemShut {NoStop}%
\bibitem [{\citenamefont {Por\'ee}\ \emph {et~al.}(2023)\citenamefont {Por\'ee}, \citenamefont {Yan}, \citenamefont {Desrochers}, \citenamefont {Petit}, \citenamefont {Lhotel}, \citenamefont {Appel}, \citenamefont {Ollivier}, \citenamefont {Kim}, \citenamefont {Nevidomskyy},\ and\ \citenamefont {Sibille}}]{poree2023fractional}%
  \BibitemOpen
  \bibfield  {author} {\bibinfo {author} {\bibfnamefont {V.}~\bibnamefont {Por\'ee}}, \bibinfo {author} {\bibfnamefont {H.}~\bibnamefont {Yan}}, \bibinfo {author} {\bibfnamefont {F.}~\bibnamefont {Desrochers}}, \bibinfo {author} {\bibfnamefont {S.}~\bibnamefont {Petit}}, \bibinfo {author} {\bibfnamefont {E.}~\bibnamefont {Lhotel}}, \bibinfo {author} {\bibfnamefont {M.}~\bibnamefont {Appel}}, \bibinfo {author} {\bibfnamefont {J.}~\bibnamefont {Ollivier}}, \bibinfo {author} {\bibfnamefont {Y.~B.}\ \bibnamefont {Kim}}, \bibinfo {author} {\bibfnamefont {A.~H.}\ \bibnamefont {Nevidomskyy}},\ and\ \bibinfo {author} {\bibfnamefont {R.}~\bibnamefont {Sibille}},\ }\href@noop {} {\  (\bibinfo {year} {2023})},\ \Eprint {https://arxiv.org/abs/2304.05452} {arXiv:2304.05452} \BibitemShut {NoStop}%
\bibitem [{\citenamefont {Li}\ and\ \citenamefont {Chen}(2017)}]{liSymmetryEnrichedTopological2017}%
  \BibitemOpen
  \bibfield  {author} {\bibinfo {author} {\bibfnamefont {Y.-D.}\ \bibnamefont {Li}}\ and\ \bibinfo {author} {\bibfnamefont {G.}~\bibnamefont {Chen}},\ }\href {https://doi.org/10.1103/PhysRevB.95.041106} {\bibfield  {journal} {\bibinfo  {journal} {Physical Review B}\ }\textbf {\bibinfo {volume} {95}},\ \bibinfo {pages} {041106(R)} (\bibinfo {year} {2017})}\BibitemShut {NoStop}%
\bibitem [{\citenamefont {Yao}\ \emph {et~al.}(2020)\citenamefont {Yao}, \citenamefont {Li},\ and\ \citenamefont {Chen}}]{yaoPyrochloreSpinLiquid2020}%
  \BibitemOpen
  \bibfield  {author} {\bibinfo {author} {\bibfnamefont {X.-P.}\ \bibnamefont {Yao}}, \bibinfo {author} {\bibfnamefont {Y.-D.}\ \bibnamefont {Li}},\ and\ \bibinfo {author} {\bibfnamefont {G.}~\bibnamefont {Chen}},\ }\href {https://doi.org/10.1103/PhysRevResearch.2.013334} {\bibfield  {journal} {\bibinfo  {journal} {Physical Review Research}\ }\textbf {\bibinfo {volume} {2}},\ \bibinfo {pages} {013334} (\bibinfo {year} {2020})}\BibitemShut {NoStop}%
\bibitem [{\citenamefont {Lantagne-Hurtubise}\ \emph {et~al.}(2017)\citenamefont {Lantagne-Hurtubise}, \citenamefont {Bhattacharjee},\ and\ \citenamefont {Moessner}}]{lantagne-hurtubiseElectricFieldControl2017}%
  \BibitemOpen
  \bibfield  {author} {\bibinfo {author} {\bibfnamefont {E.}~\bibnamefont {Lantagne-Hurtubise}}, \bibinfo {author} {\bibfnamefont {S.}~\bibnamefont {Bhattacharjee}},\ and\ \bibinfo {author} {\bibfnamefont {R.}~\bibnamefont {Moessner}},\ }\href {https://doi.org/10.1103/PhysRevB.96.125145} {\bibfield  {journal} {\bibinfo  {journal} {Physical Review B}\ }\textbf {\bibinfo {volume} {96}},\ \bibinfo {pages} {125145} (\bibinfo {year} {2017})}\BibitemShut {NoStop}%
\bibitem [{\citenamefont {Mandal}(2019)}]{mandalElectricFieldResponse2019}%
  \BibitemOpen
  \bibfield  {author} {\bibinfo {author} {\bibfnamefont {I.}~\bibnamefont {Mandal}},\ }\href {https://doi.org/10.1140/epjb/e2019-100215-4} {\bibfield  {journal} {\bibinfo  {journal} {The European Physical Journal B}\ }\textbf {\bibinfo {volume} {92}},\ \bibinfo {pages} {187} (\bibinfo {year} {2019})}\BibitemShut {NoStop}%
\bibitem [{\citenamefont {Patri}\ \emph {et~al.}(2020)\citenamefont {Patri}, \citenamefont {Hosoi},\ and\ \citenamefont {Kim}}]{PatriPhysRevResearch}%
  \BibitemOpen
  \bibfield  {author} {\bibinfo {author} {\bibfnamefont {A.~S.}\ \bibnamefont {Patri}}, \bibinfo {author} {\bibfnamefont {M.}~\bibnamefont {Hosoi}},\ and\ \bibinfo {author} {\bibfnamefont {Y.~B.}\ \bibnamefont {Kim}},\ }\href {https://doi.org/10.1103/PhysRevResearch.2.023253} {\bibfield  {journal} {\bibinfo  {journal} {Phys. Rev. Research}\ }\textbf {\bibinfo {volume} {2}},\ \bibinfo {pages} {023253} (\bibinfo {year} {2020})}\BibitemShut {NoStop}%
\bibitem [{SM()}]{SM}%
  \BibitemOpen
  \href@noop {} {}\bibinfo {note} {See Supplemental Material at \url{Placeholder URL} for details of perturbation theory, numerics and large-$S$ formalism.}\BibitemShut {Stop}%
\bibitem [{\citenamefont {Villain}(1979)}]{villainInsulatingSpinGlasses1979}%
  \BibitemOpen
  \bibfield  {author} {\bibinfo {author} {\bibfnamefont {J.}~\bibnamefont {Villain}},\ }\href {https://doi.org/10.1007/BF01325811} {\bibfield  {journal} {\bibinfo  {journal} {Zeitschrift f\"ur Physik B Condensed Matter and Quanta}\ }\textbf {\bibinfo {volume} {33}},\ \bibinfo {pages} {31} (\bibinfo {year} {1979})}\BibitemShut {NoStop}%
\bibitem [{\citenamefont {Kwasigroch}\ \emph {et~al.}(2017)\citenamefont {Kwasigroch}, \citenamefont {Dou\c{c}ot},\ and\ \citenamefont {Castelnovo}}]{kwasigrochSemiclassicalApproachQuantum2017}%
  \BibitemOpen
  \bibfield  {author} {\bibinfo {author} {\bibfnamefont {M.~P.}\ \bibnamefont {Kwasigroch}}, \bibinfo {author} {\bibfnamefont {B.}~\bibnamefont {Dou\c{c}ot}},\ and\ \bibinfo {author} {\bibfnamefont {C.}~\bibnamefont {Castelnovo}},\ }\href {https://doi.org/10.1103/PhysRevB.95.134439} {\bibfield  {journal} {\bibinfo  {journal} {Physical Review B}\ }\textbf {\bibinfo {volume} {95}},\ \bibinfo {pages} {134439} (\bibinfo {year} {2017})}\BibitemShut {NoStop}%
\bibitem [{\citenamefont {Benton}\ \emph {et~al.}(2016)\citenamefont {Benton}, \citenamefont {Sikora},\ and\ \citenamefont {Shannon}}]{Benton2016PhysRev}%
  \BibitemOpen
  \bibfield  {author} {\bibinfo {author} {\bibfnamefont {O.}~\bibnamefont {Benton}}, \bibinfo {author} {\bibfnamefont {O.}~\bibnamefont {Sikora}},\ and\ \bibinfo {author} {\bibfnamefont {N.}~\bibnamefont {Shannon}},\ }\href {https://doi.org/10.1103/PhysRevB.93.125143} {\bibfield  {journal} {\bibinfo  {journal} {Phys. Rev. B}\ }\textbf {\bibinfo {volume} {93}},\ \bibinfo {pages} {125143} (\bibinfo {year} {2016})}\BibitemShut {NoStop}%
\bibitem [{\citenamefont {Chen}(2017{\natexlab{b}})}]{chenSpectralPeriodicitySpinon2017}%
  \BibitemOpen
  \bibfield  {author} {\bibinfo {author} {\bibfnamefont {G.}~\bibnamefont {Chen}},\ }\href {https://doi.org/10.1103/PhysRevB.96.085136} {\bibfield  {journal} {\bibinfo  {journal} {Physical Review B}\ }\textbf {\bibinfo {volume} {96}},\ \bibinfo {pages} {085136} (\bibinfo {year} {2017}{\natexlab{b}})}\BibitemShut {NoStop}%
\bibitem [{\citenamefont {Desrochers}\ \emph {et~al.}(2023)\citenamefont {Desrochers}, \citenamefont {Chern},\ and\ \citenamefont {Kim}}]{desrochersSymmetryFractionalizationGauge2023}%
  \BibitemOpen
  \bibfield  {author} {\bibinfo {author} {\bibfnamefont {F.}~\bibnamefont {Desrochers}}, \bibinfo {author} {\bibfnamefont {L.~E.}\ \bibnamefont {Chern}},\ and\ \bibinfo {author} {\bibfnamefont {Y.~B.}\ \bibnamefont {Kim}},\ }\href {https://doi.org/10.1103/PhysRevB.107.064404} {\bibfield  {journal} {\bibinfo  {journal} {Physical Review B}\ }\textbf {\bibinfo {volume} {107}},\ \bibinfo {pages} {064404} (\bibinfo {year} {2023})}\BibitemShut {NoStop}%
\bibitem [{\citenamefont {Zhou}\ \emph {et~al.}(2024)\citenamefont {Zhou}, \citenamefont {Desrochers},\ and\ \citenamefont {Kim}}]{zhou2024arxiv}%
  \BibitemOpen
  \bibfield  {author} {\bibinfo {author} {\bibfnamefont {Z.}~\bibnamefont {Zhou}}, \bibinfo {author} {\bibfnamefont {F.}~\bibnamefont {Desrochers}},\ and\ \bibinfo {author} {\bibfnamefont {Y.~B.}\ \bibnamefont {Kim}},\ }\href {https://arxiv.org/abs/2406.18650} {\bibinfo {title} {Magnetic field response of dipolar-octupolar quantum spin ice}} (\bibinfo {year} {2024}),\ \Eprint {https://arxiv.org/abs/2406.18650} {arXiv:2406.18650 [cond-mat.str-el]} \BibitemShut {NoStop}%
\bibitem [{\citenamefont {Park}\ \emph {et~al.}(2020)\citenamefont {Park}, \citenamefont {Kim},\ and\ \citenamefont {Lee}}]{parkGeometricSuperconductivity3D2020}%
  \BibitemOpen
  \bibfield  {author} {\bibinfo {author} {\bibfnamefont {M.~J.}\ \bibnamefont {Park}}, \bibinfo {author} {\bibfnamefont {Y.~B.}\ \bibnamefont {Kim}},\ and\ \bibinfo {author} {\bibfnamefont {S.}~\bibnamefont {Lee}},\ }\href {https://doi.org/10.48550/arXiv.2007.16205} {\bibinfo {title} {Geometric {{Superconductivity}} in {{3D Hofstadter Butterfly}}}} (\bibinfo {year} {2020}),\ \Eprint {https://arxiv.org/abs/2007.16205} {2007.16205} \BibitemShut {NoStop}%
\bibitem [{\citenamefont {Hofstadter}(1976)}]{hofstadterEnergyLevelsWave1976}%
  \BibitemOpen
  \bibfield  {author} {\bibinfo {author} {\bibfnamefont {D.~R.}\ \bibnamefont {Hofstadter}},\ }\href {https://doi.org/10.1103/PhysRevB.14.2239} {\bibfield  {journal} {\bibinfo  {journal} {Physical Review B}\ }\textbf {\bibinfo {volume} {14}},\ \bibinfo {pages} {2239} (\bibinfo {year} {1976})}\BibitemShut {NoStop}%
\bibitem [{\citenamefont {Yang}\ \emph {et~al.}(2020)\citenamefont {Yang}, \citenamefont {Kim},\ and\ \citenamefont {Lee}}]{yangMagneticFieldThermal2020}%
  \BibitemOpen
  \bibfield  {author} {\bibinfo {author} {\bibfnamefont {H.-J.}\ \bibnamefont {Yang}}, \bibinfo {author} {\bibfnamefont {H.~S.}\ \bibnamefont {Kim}},\ and\ \bibinfo {author} {\bibfnamefont {S.~B.}\ \bibnamefont {Lee}},\ }\href {https://doi.org/10.1103/PhysRevB.102.060405} {\bibfield  {journal} {\bibinfo  {journal} {Physical Review B}\ }\textbf {\bibinfo {volume} {102}},\ \bibinfo {pages} {060405(R)} (\bibinfo {year} {2020})}\BibitemShut {NoStop}%
\bibitem [{Note1()}]{Note1}%
  \BibitemOpen
  \bibinfo {note} {Note that, due to the 3D non-collinear nature of the local easy axes, it is not possible to apply a purely transverse external field.}\BibitemShut {Stop}%
\bibitem [{\citenamefont {Laumann}\ and\ \citenamefont {Moessner}(2023)}]{laumannHybridDyonsInverted2023}%
  \BibitemOpen
  \bibfield  {author} {\bibinfo {author} {\bibfnamefont {C.~R.}\ \bibnamefont {Laumann}}\ and\ \bibinfo {author} {\bibfnamefont {R.}~\bibnamefont {Moessner}},\ }\href {http://arxiv.org/abs/2302.06635} {\  (\bibinfo {year} {2023})},\ \Eprint {https://arxiv.org/abs/arXiv:2302.06635} {arXiv:2302.06635} \BibitemShut {NoStop}%
\bibitem [{\citenamefont {Yahne}\ \emph {et~al.}(2024)\citenamefont {Yahne}, \citenamefont {Placke}, \citenamefont {Sch\"afer}, \citenamefont {Benton}, \citenamefont {Moessner}, \citenamefont {Powell}, \citenamefont {Kolis}, \citenamefont {Pasco}, \citenamefont {May}, \citenamefont {Frontzek}, \citenamefont {Smith}, \citenamefont {Gaulin}, \citenamefont {Calder},\ and\ \citenamefont {Ross}}]{yahne2022dipolar}%
  \BibitemOpen
  \bibfield  {author} {\bibinfo {author} {\bibfnamefont {D.~R.}\ \bibnamefont {Yahne}}, \bibinfo {author} {\bibfnamefont {B.}~\bibnamefont {Placke}}, \bibinfo {author} {\bibfnamefont {R.}~\bibnamefont {Sch\"afer}}, \bibinfo {author} {\bibfnamefont {O.}~\bibnamefont {Benton}}, \bibinfo {author} {\bibfnamefont {R.}~\bibnamefont {Moessner}}, \bibinfo {author} {\bibfnamefont {M.}~\bibnamefont {Powell}}, \bibinfo {author} {\bibfnamefont {J.~W.}\ \bibnamefont {Kolis}}, \bibinfo {author} {\bibfnamefont {C.~M.}\ \bibnamefont {Pasco}}, \bibinfo {author} {\bibfnamefont {A.~F.}\ \bibnamefont {May}}, \bibinfo {author} {\bibfnamefont {M.~D.}\ \bibnamefont {Frontzek}}, \bibinfo {author} {\bibfnamefont {E.~M.}\ \bibnamefont {Smith}}, \bibinfo {author} {\bibfnamefont {B.~D.}\ \bibnamefont {Gaulin}}, \bibinfo {author} {\bibfnamefont {S.}~\bibnamefont {Calder}},\ and\ \bibinfo {author} {\bibfnamefont {K.~A.}\ \bibnamefont {Ross}},\ }\href {https://doi.org/10.1103/PhysRevX.14.011005} {\bibfield  {journal} {\bibinfo  {journal} {Phys. Rev. X}\ }\textbf {\bibinfo {volume} {14}},\ \bibinfo {pages} {011005} (\bibinfo {year} {2024})}\BibitemShut {NoStop}%
\bibitem [{\citenamefont {Smith}\ \emph {et~al.}(2023)\citenamefont {Smith}, \citenamefont {Dudemaine}, \citenamefont {Placke}, \citenamefont {Sch\"afer}, \citenamefont {Yahne}, \citenamefont {DeLazzer}, \citenamefont {Fitterman}, \citenamefont {Beare}, \citenamefont {Gaudet}, \citenamefont {Buhariwalla}, \citenamefont {Podlesnyak}, \citenamefont {Xu}, \citenamefont {Clancy}, \citenamefont {Movshovich}, \citenamefont {Luke}, \citenamefont {Ross}, \citenamefont {Moessner}, \citenamefont {Benton}, \citenamefont {Bianchi},\ and\ \citenamefont {Gaulin}}]{Smith2023PhysRevB}%
  \BibitemOpen
  \bibfield  {author} {\bibinfo {author} {\bibfnamefont {E.~M.}\ \bibnamefont {Smith}}, \bibinfo {author} {\bibfnamefont {J.}~\bibnamefont {Dudemaine}}, \bibinfo {author} {\bibfnamefont {B.}~\bibnamefont {Placke}}, \bibinfo {author} {\bibfnamefont {R.}~\bibnamefont {Sch\"afer}}, \bibinfo {author} {\bibfnamefont {D.~R.}\ \bibnamefont {Yahne}}, \bibinfo {author} {\bibfnamefont {T.}~\bibnamefont {DeLazzer}}, \bibinfo {author} {\bibfnamefont {A.}~\bibnamefont {Fitterman}}, \bibinfo {author} {\bibfnamefont {J.}~\bibnamefont {Beare}}, \bibinfo {author} {\bibfnamefont {J.}~\bibnamefont {Gaudet}}, \bibinfo {author} {\bibfnamefont {C.~R.~C.}\ \bibnamefont {Buhariwalla}}, \bibinfo {author} {\bibfnamefont {A.}~\bibnamefont {Podlesnyak}}, \bibinfo {author} {\bibfnamefont {G.}~\bibnamefont {Xu}}, \bibinfo {author} {\bibfnamefont {J.~P.}\ \bibnamefont {Clancy}}, \bibinfo {author} {\bibfnamefont {R.}~\bibnamefont {Movshovich}}, \bibinfo {author} {\bibfnamefont {G.~M.}\ \bibnamefont {Luke}}, \bibinfo {author} {\bibfnamefont {K.~A.}\ \bibnamefont {Ross}}, \bibinfo {author} {\bibfnamefont {R.}~\bibnamefont {Moessner}}, \bibinfo {author} {\bibfnamefont {O.}~\bibnamefont {Benton}}, \bibinfo {author} {\bibfnamefont {A.~D.}\ \bibnamefont {Bianchi}},\ and\ \bibinfo {author} {\bibfnamefont {B.~D.}\ \bibnamefont {Gaulin}},\ }\href {https://doi.org/10.1103/PhysRevB.108.054438} {\bibfield  {journal} {\bibinfo  {journal} {Phys. Rev. B}\ }\textbf {\bibinfo {volume} {108}},\ \bibinfo {pages} {054438} (\bibinfo {year} {2023})}\BibitemShut {NoStop}%
\bibitem [{\citenamefont {Bhardwaj}\ \emph {et~al.}(2022)\citenamefont {Bhardwaj}, \citenamefont {Zhang}, \citenamefont {Yan}, \citenamefont {Moessner}, \citenamefont {Nevidomskyy},\ and\ \citenamefont {Changlani}}]{Bhardwaj2022NPJQM}%
  \BibitemOpen
  \bibfield  {author} {\bibinfo {author} {\bibfnamefont {A.}~\bibnamefont {Bhardwaj}}, \bibinfo {author} {\bibfnamefont {S.}~\bibnamefont {Zhang}}, \bibinfo {author} {\bibfnamefont {H.}~\bibnamefont {Yan}}, \bibinfo {author} {\bibfnamefont {R.}~\bibnamefont {Moessner}}, \bibinfo {author} {\bibfnamefont {A.~H.}\ \bibnamefont {Nevidomskyy}},\ and\ \bibinfo {author} {\bibfnamefont {H.~J.}\ \bibnamefont {Changlani}},\ }\href {http://dx.doi.org/10.1038/s41535-022-00458-2} {\bibfield  {journal} {\bibinfo  {journal} {npj Quantum Materials}\ }\textbf {\bibinfo {volume} {7}},\ \bibinfo {pages} {51} (\bibinfo {year} {2022})}\BibitemShut {NoStop}%
\end{thebibliography}%

\clearpage
\setcounter{equation}{0}
\setcounter{figure}{0}
\setcounter{table}{0}
\makeatletter
\renewcommand{\theequation}{S\arabic{equation}}
\renewcommand{\thefigure}{S\arabic{figure}}
\renewcommand{\bibnumfmt}[1]{[#1]}
\renewcommand{\citenumfont}[1]{#1}

\begin{widetext}
\begin{center}
{	\Large{\textbf{Supplementary Materials for ``\thistitle"}} } 

\vspace{5mm}
 Alaric Sanders, Han Yan, Claudio Castelnovo and Andriy H. Nevidomskyy 
 
\end{center}
\end{widetext}

\section{Perturbation Theory}
We begin with the Hamiltonian $\mathcal{H}'$ in Eq.~(2) of the main text:
\begin{align}
\mathcal{H}' &= 
\sum_{\langle i j \rangle} 
\frac{J_\pm}{2} (\sigma_i^+\sigma_j^- + \sigma_i^-\sigma_j^+ )
-\sum_i \frac{\vb*{\hat{z}}_i \cdot \vb*{B}}{2}
(\sigma_i^+ + \sigma_i^-) 
\end{align}
By performing a perturbative expansion with respect to  $\mathcal{H}^{\prime}$, relative to the Hamiltonian $\mathcal{H}_I=\sum_{\langle i j \rangle}  J_y \sigma_i^z \sigma_j^z$ enforcing the spin-ice rules,  we obtain an effective Hamiltonian of the general form
\[
\mathcal{H}_{\text {eff}} = (1-\mathcal{P})
\bigg[
  -\mathcal{H}^{\prime} \frac{\mathcal{P}}{\mathcal{H}_I} \mathcal{H}^{\prime}
  +\mathcal{H}^{\prime} \frac{\mathcal{P}}{\mathcal{H}_I} \mathcal{H}^{\prime} \frac{\mathcal{P}}{\mathcal{H}_I} \mathcal{H}^{\prime}  
+ \dots 
\bigg] (1-\mathcal{P}) 
\, , 
\label{eqS:ham_expansion}
\]
where $1-\mathcal{P}$ is the projection operator onto the ice states. The only non-vanishing terms in the above expansion are therefore those that map spin ice configurations to spin ice configurations, with the affected spins forming closed loops (or trivial self-retracing paths).

It is useful to split $\mathcal{H}'$ into two contributions, the XX part $\hxxz = \sum_{\langle ij \rangle }\frac{J_\pm}{2} (\sigma^+_i\sigma^-_j + \sigma^-_i\sigma^+_j)$ common to all QSI models and the Zeeman term $\hz = - \sum_i \frac{\vb*{z}_i \cdot \vb*{B}}{2} \sigma^x_i$ unique to DO-QSI. 

The ice rule constraint immediately excludes any contribution at first order.
At second order, we have two non-vanishing contributions. From $\hxxz \frac{\mathcal{P}}{\mathcal{H}_I}\hxxz$, we obtain only a term proportional to $J_\pm^2 \{\sigma_i^+\sigma_j^-, \sigma_i^-\sigma_j^+\}$ arising from two factors of $\sigma^+\sigma^-$ occurring on the same pyrochlore bond. This is a term from standard spin ice~\cite{Hermele_Fisher_Balents}, which counts the number of nearest-neighbour, oppositely aligned spins. Within the ice manifold, each tetrahedron has exactly six antialigned nearest-neighbour spins, and this term is therefore a trivial constant.
The second contribution, arising from $\hz \frac{\mathcal{P}}{\mathcal{H}_I}\hz$, is proportional to $B^2 \{\sigma_i^+, \sigma_i^-\}$ and so 
constant.

At third order, we find non-vanishing terms from (I) $\hxxz  \frac{\mathcal{P}}{\mathcal{H}_I}\hxxz \frac{\mathcal{P}}{\mathcal{H}_I}\hxxz$ and (II) permutations of 
$\hz  \frac{\mathcal{P}}{\mathcal{H}_I}\hz \frac{\mathcal{P}}{\mathcal{H}_I}\hxxz$. 
From (I), we obtain further trivial constant terms and the well known hexagon-flip term \mbox{$\frac{3 J_\pm^3}{2 J_y^2} (\mathcal{O} + \mathcal{O}^\dagger)$}, while a short calculation reveals (II) to be a constant.

\begin{figure}
\includegraphics{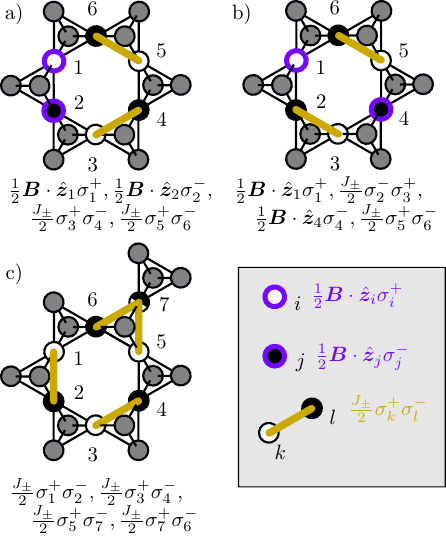}
	\caption{
		\label{fig:ringflip_channels}
		Hexagon-flip processes occurring in the perturbative expansion. Following the convention of Refs.~\onlinecite{Hermele_Fisher_Balents,sandersVisonCrystalQuantum2023}, we represent spins as hard-core bosons, with down (up) spins written as empty (filled) circles, all in the $S^y$ basis. Gray circles are spectator spins that are decoupled from the process. The two leading order processes responsible for field-mediated hexagon-flip terms are show in panels a) and b). Panel c) shows the fourth order pure-XX correction to the hexagon-flip Hamiltonian, note that spin 7 is flipped twice. See Ref.~\onlinecite{sandersVisonCrystalQuantum2023} for further remarks on the combinatorics associated with these diagrams.
	}
\end{figure}

The leading order non-trivial effect of the magnetic field comes at fourth order, from processes similar to those depicted in Fig.~\ref{Fig_lattice} in the main text. Two channels contribute to this process, corresponding to different arrangements of the field-mediated single-spin flips, see Fig.~\ref{fig:ringflip_channels}a,b). 

It may be show from these considerations that the effective Hamiltonian at order
$J_\pm^3/J_y^2$ and $B^2 J_\pm^2/J_y^3$ is of the form
\begin{equation}
\mathcal{H}_\text{eff}
= \sum_{p \in \{\text{all } \hexagon \}} 
\Jring_p 
\left[ 
  \mathcal{O}_{p} + \mathcal{O}_{p}^\dagger 
\right]\ .
\label{eqS:eff_hamiltonian}
\end{equation}

In practice, we calculate the values of $g_p$ on each type of hexagon by six site exact diagonalisation. Starting from the fourth order series of Eq.~\eqref{eqS:ham_expansion}, we project $\mathcal{H}_I,\ \mathcal{H}',$ and $\mathcal{P}$ on to the basis of all possible hexagon states and so recast them as $2^6$ by $2^6$ matrices.
The coefficient $g_p$ given in Eq.~\eqref{EQN_general_Jring} of the main text then corresponds to the matrix element of $\mathcal{H}_{\rm eff}$ connecting
the state of clockwise pointing spins to the state of counter-clockwise pointing spins. 
This calculation includes all the relevant spinon creation, propagation and annihilation processes implicitly, automatically including all relevant combinatorial factors.
 
\begin{figure}[t!]
\includegraphics[width=\columnwidth]{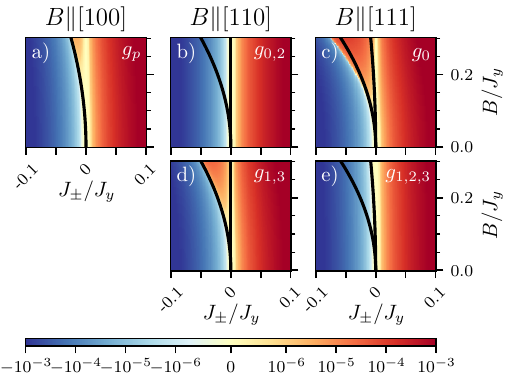}
\caption{
\label{Supp_Fig_Jring_all}
Raw effective hexagon-flip operator coefficient strength from the leading-order expansion, Eq.~\eqref{EQN_general_Jring}. The color scale is in units of $J_y$. Note that $J_\pm = 0$ marks the line on which all standard QSI hexagon-flips vanish, i.e., the classical limit.
}
\end{figure}

Other higher order contributions, such as the pure XX contribution at fourth order shown in Fig.~\ref{fig:ringflip_channels}c) can be generated when one includes higher order perturbations and more neighbouring spins, but are neglected in our work since they do not alter the qualitative features of the phase diagram in Fig.~\ref{Supp_Fig_Jring_all}. 

\section{Semiclassical simulations}
We studied the effective Hamiltonian in Eq.~\eqref{eqS:eff_hamiltonian} using the semiclassical method detailed in Refs.~\onlinecite{szaboSeeingLightVison2019a,sandersVisonCrystalQuantum2023}, which effectively produces a zeroth-order approximation to the imaginary-time path integral of the system. 

In this approach, we simulate finite temperature dynamics by the following procedure:
\begin{enumerate}
    \item Using a classical Metropolis algorithm, we generate an ensemble of equilibrium configurations at the desired temperature.
    \item For each of these configurations, we perform classical Landau--Lifshitz time evolution. 
    \item Finally, we average any observables over the ensemble of thermal configurations.
\end{enumerate}
We simulate a periodic cubic supercell of $12^3 \times 16 = 27\,648$ spins. Starting from a temperature of $T = 10$ (in units of $J_\pm^3/J_y^2$), we anneal to the final temperature via $256$ logarithmically-spaced steps, using Monte Carlo acceptance rates to monitor convergence. We average the time evolution dynamics over an ensemble of $32$ thermal configurations. In all other respects, the simulation parameters are the same as those used in Ref.~\onlinecite{sandersVisonCrystalQuantum2023}.

\section{Relevance to DO-QSI candidate materials}
In Table~\ref{tab:materials} we summarise for convenience some relevant details of the cerium-based pyrochlore materials studied in the literature to date. 
\begin{table}[ht!]
    \begin{tabular}{c|ccccc}
        Material &$ J_y $&$ J_x$&$ J_z $&$ J_{xz} $&$\begin{matrix}
            \text{single}\\
            \text{crystal}
        \end{matrix}$ \\\hline
         $\mathrm{Ce}_2\mathrm{Hf}_2\mathrm{O}_7$~\cite{poree2023dipolaroctupolar} %
         &$0.044 $&$0.011$&$ 0.016 $&$-0.002 $&  Yes\footnote{private communication} \\ 
         $\mathrm{Ce}_2\mathrm{Hf}_2\mathrm{O}_7$ \cite{poree2023dipolaroctupolar} &$0.022$&$0.046$&$ 0.011 $&$-0.001 $& Yes \\ 
         $\mathrm{Ce}_2\mathrm{Sn}_2\mathrm{O}_7$ \cite{yahne2022dipolar} &$-0.001$ &$0.045$ &$-0.012$ &$0$ & No \\
         $\mathrm{Ce}_2\mathrm{Sn}_2\mathrm{O}_7$ \cite{Sibille2020NatPhys,poree2023fractional} &$0.069$ &$0.034$ &$0.034$ &$0$ & No \\
         $\mathrm{Ce}_2\mathrm{Zr}_2\mathrm{O}_7$ \cite{Smith2023PhysRevB} &$ 0.063 $&$0.062 $&$ 0.011 $&$ 0 $ & Yes\\ 
         $\mathrm{Ce}_2\mathrm{Zr}_2\mathrm{O}_7$ \cite{Bhardwaj2022NPJQM} &$ 0.069 $&$0.068 $&$ 0.013 $&$ 0 $ & Yes\\ 
    \end{tabular}
    \caption{\label{tab:materials}
        Estimated coupling parameters for DO-QSI candidates, quoted in meV.
    }
\end{table}

\section{Large-$S$ photon dispersion}
\label{sec:photon-dispersion}
The spin Hamiltonian of Eq.~\eqref{eqS:eff_hamiltonian} admits an analytical solution via the Villain expansion given in 
the main text. To quadratic order in spin fluctuations, we obtain~\cite{kwasigrochSemiclassicalApproachQuantum2017,BentonPhysRevB}
\[
\mathcal{H}_{\mathrm{large-}S} \simeq \sum_{j \in \{\rm{Spins} \}} U_{\alpha(j)} (\sigma^z_j(t))^2 - \sum_{p \in \{\hexagon \} }g_{\mu(p)}\cos(\Phi_p(t))
\, , 
\]
where $\mu(p) \in \{0,1,2,3\}$ denotes the type of hexagon, $\alpha(j) \in \{0,1,2,3\}$ denotes the sublattice of spin $j$, and $U_{\alpha(j)} = 2 \sum_{\mu \neq \alpha} g_\mu$ represents a sum over the six hexagon-flip coefficients bordering spin $j$. 

Noting that in the ice-state manifold, $\langle \sigma^z \rangle = 0$, we first statically extremise this Hamiltonian and obtain the zero-temperature flux configuration $\Phi^{\rm GS}_p$. We find that this always results in one of the four possible ground states: 0-flux, $\pi$-flux, $\pi\pi00$ flux or frustrated flux (FF) state. We then expand in fluctuations, setting $\Phi_p(t) = \Phi^{\rm GS}_p + \delta \Phi_p(t)$ and truncating the series at quadratic order in $\delta\Phi$. Working in the radiation gauge, $\sigma^z$ is identified with $\partial_t a_j$ and the normal modes found in the usual way.

In order to provide a concrete expression for the quadratic modes, we are now forced to change index convention. The spins $j$ of the pyrochlore lattice may be uniquely indexed in terms of an FCC lattice site $\vb*{r}_\wedge$, corresponding to the centre of an up-pointing tetrahedron, and a sublattice index $\alpha \in \{0,1,2,3\}$. We find that the Fourier modes of $\sigma^z$ and $a$, 
\begin{align*}
    a_{\vb*{r_\wedge}+\boldsymbol{e}_{\alpha}/2}(t) &= \sqrt{\frac{2}{N_{t}}} \sum_{\boldsymbol{k}} e^{i \boldsymbol{k} \cdot (\vb*{r_\wedge}+ \boldsymbol{e}_\alpha/2)}a_{\alpha}(\boldsymbol{k},t)\\
\sigma^z_{\vb*{r_\wedge}+\boldsymbol{e}_{\alpha}/2}(t)&= \sqrt{\frac{2}{N_{t}}}\sum_{\boldsymbol{k}}e^{i \boldsymbol{k} \cdot (\vb*{r_\wedge}+ \boldsymbol{e}_\alpha/2)}\sigma^z_{\alpha}(\boldsymbol{k},t)
\, , 
\end{align*}
evolve as
\begin{align*}
\partial_t a_{\alpha}( \vb*{k},t) &= \sum_{\mu\neq \alpha} 2g_\mu \sigma^z_{\alpha} (-\vb*{k},t)\\
\partial_t \sigma^z_{\alpha}(\vb*{k},t) &=
- \sum_\nu g_\nu \sin(\vb*{h}_{\alpha\nu} \cdot \vb*{k})\sin(\vb*{h}_{\beta \nu} \cdot \vb*{k}) a_{\beta}(-\vb*{k}, t)\\
\vb*{h}_{\mu\nu} &= \frac{a_0}{\sqrt{8}}\frac{\boldsymbol{e}_{\mu} \times \boldsymbol{e}_{\lambda}}{\|\boldsymbol{e}_{\mu}\times \boldsymbol{e}_\lambda\|}
\, , 
\end{align*}
where $\boldsymbol{e}_\alpha$ denotes the vector joining the centre of an `up' tetrahedron to the spin on sublattice $\alpha$, and $a_0$ denotes the cubic lattice parameter. 

\section{Photon up-conversion in the FF case}
As Fig.~\ref{fig:analytics+numerics}d in the main text shows, the FF phase is distinct from the non-frustrated cases in that a structured continuum of excitations with several step-like features appears.  
In optics, this phenomenon is known as three-photon up-conversion.
Fundamentally, it arises from the expansion of $\cos \Phi_p$ about its zero-temperature expectation value: the series expansions of $\cos(0 + \delta \Phi)$ and $\cos(\pi + \delta \Phi)$ contain only even powers of $\delta \Phi$, while no such symmetry applies in the FF case. In the large-$S$ formalism, any beyond-quadratic terms in the expansion are interaction vertices - it is only in the FF phase that the $\delta\Phi^3$ vertex survives.
It can readily be shown~\cite{sandersVisonCrystalQuantum2023} that pairs of contracted three-photon vertices have the same order in renormalization group as the Hartree-Fock corrections in large-$S$ theory, suggesting that these effects will be of similar importance for spin-$1/2$ moments (note that Hartree-Fock vertices are believed to correct the speed of light by a factor of $2.7$~\cite{kwasigrochSemiclassicalApproachQuantum2017}).

\section{Dualities of the model}
The semiclassical analysis we presented amounts to expanding the emergent magnetic field flux \mbox{$b(t) = \Phi_0 + \delta b(t)$} to quadratic order in the field fluctuations $\delta b(t)$, where $\Phi_0$ is the static flux background as depicted in  Figs.~\ref{Fig_Jring_all} and~\ref{fig:magnetic_field_tuning} of the main text. This allows us to map the gauge dynamics of the $U(1)_\pi$ and $\pi\pi 00$ phases onto the $U(1)_0$ spin ice phase. Here we show that this mapping also works at the level of the effective spin Hamiltonian, Eq.~\eqref{eq:eff_hamiltonian}.

We define the unitary operator 
\begin{align}
U[a] = \prod_{j} \exp(i a_j \sigma^z_j)
\, , 
\end{align}
where the index $j$ runs over all spins, and $a_j \in [0,2\pi)$.
It can be shown that $U[a] \mathcal{O}_{p} U[a]^\dagger = e^{i \nabla_{p} \times a} \mathcal{O}_{p}$. The lattice curl $\nabla_{p}\times = a_1-a_2+a_3-a_4+a_5-a_6$ is the alternating sum of emergent gauge fields $a$. In the large-$S$ limit, one may interpret this as the statement that emergent magnetic fields may be added together, with the caveat of $2\pi$ periodicity. 

By `subtracting $\pi$' from two of the four sublattices, one can map the $\pi\pi00$ phase to $U(1)_0$. Given a gauge field configuration $a_{\pi\pi00}$ generating a $\pi\pi00$ state, it follows that
\begin{equation}
    \begin{split}
        U[a_{\pi\pi00}]\mathcal{H}_{\text {eff}}[g_0, g_1, g_2, g_3]&U[a_{\pi\pi00}]^\dagger 
        \\
        & = \mathcal{H}_{\text {eff}}[-g_0, -g_1, g_2, g_3]
        \, . 
    \end{split}
\end{equation} 
The matter-free dynamics of all anisotropic $\Jring$ models is therefore symmetric with respect to flipping the signs of any two $\Jring$ values (though one may expect the symmetries of the spinons to fractionalize due to the enlarged gauge unit cell).

The frustrated-flux phase, by contrast, is not dual to $U(1)_0$, as the frustration of the emergent fluxes cannot be subtracted away. For concreteness, let $a_{\rm FF}$ generate the flux configuration $[\pi, \pi/3, \pi/3, \pi/3]$, as an example. Then, letting $\mu(p)$ denote the sublattice of hexagon ${p}$, we find 
\begin{equation}
    \begin{split}
         &U[a_{\rm FF}] H[g_0, g_1, g_2, g_3] U[a_{\rm FF}]^\dagger \\
    &= 
    H\left[-g_0, \frac{g_1}{2}, \frac{g_2}{2}, \frac{g_3}{2}\right] + \sum_{p, \mu(p)\neq 0} i g_{p} \frac{\sqrt{3}}{2} (\mathcal{O}_{p} - \mathcal{O}^\dagger_{p})   
    \, . 
    \end{split}
\end{equation}
This highly unusual complex hexagon-flip term has previously been identified in spin-$1/2$ QSI on the breathing pyrochlore lattice~\cite{sandersVisonCrystalQuantum2023}. It is associated with the breaking of a) inversion symmetry and b) charge-conjugation symmetry, both of which are facilitated here by a [111] magnetic field. The finding represents the first time such a coupling has been suggested in the pyrochlore QSI literature, to the best of our knowledge.

Following Ref.~\onlinecite{sandersVisonCrystalQuantum2023}, we recognise that these terms are decoupled from the fluctuating background to leading order, manifesting instead as the three-photon terms. 

\begin{figure*}[ht]
	\includegraphics[width=\textwidth]{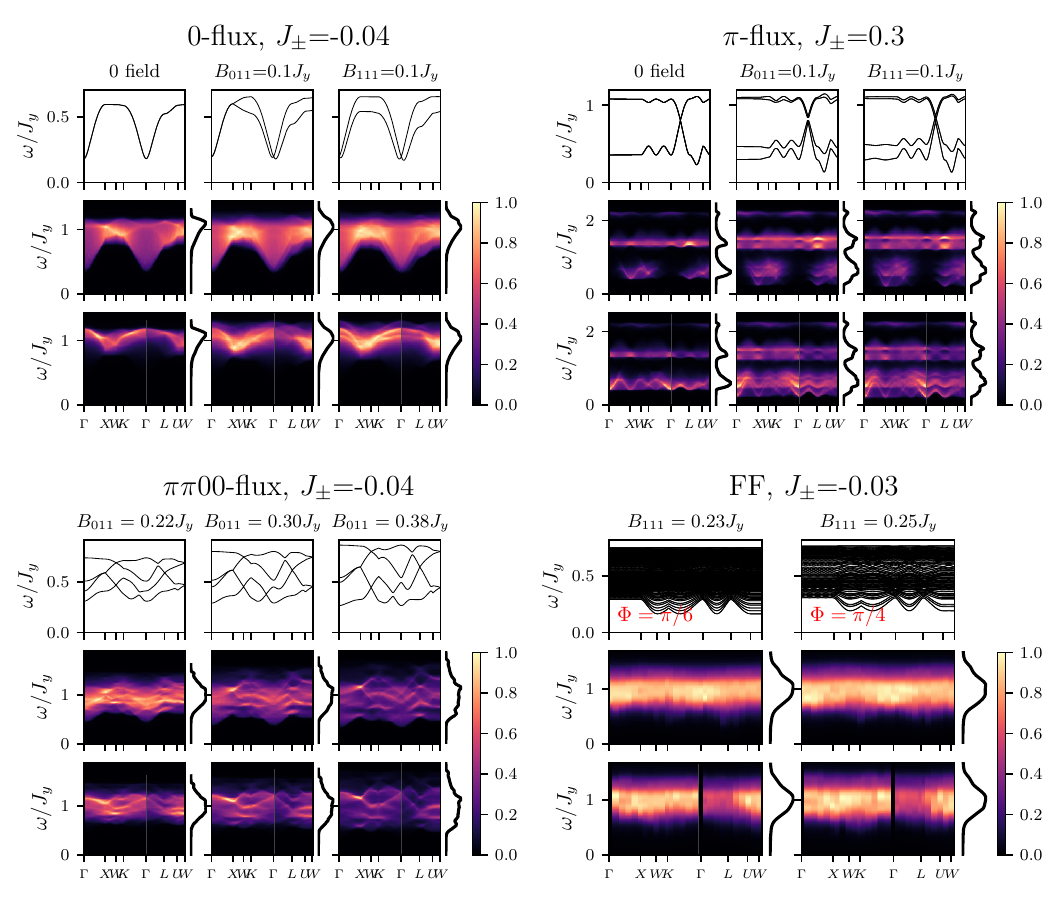}
	\caption{The four panels, each for a different DO-QSI phase, show spinon dispersions (top row), dynamical spin correlations $\langle \sigma^+(\vb*{q},\omega)\sigma^-(-\vb*{q}) \rangle$ (middle row) and magnetic structure factors $\mathcal{S}(\mathbf{q},\omega)$ (bottom row). Curves on the vertical axes of the colour plots correspond
	to Brillouin-zone integrated intensities. Red annotations in the FF panel correspond to the elementary flux $\Phi$ from Eq.~\eqref{eq:landau_gauge}. Note differing $\omega$ axis scales between panels.
	}
	\label{fig:grand-unified}
\end{figure*}

\section{Spinon continuum from gauge mean field theory}
\as{
Departing from our discussion of matter-free gauge dynamics, we consider the spinon sector of the eQED theory. %
We employ the gauge mean field theory (GMFT) formalism of Refs.~\onlinecite{desrochersSpectroscopicSignaturesFractionalization2023} and~\onlinecite{savaryCoulombicQuantumLiquids2012}. In this scheme, we make the approximation that the spinons couple to the photon field only via a static mean-field flux background, allowing one to transform the spin Hamiltonian into a quadratic tight-binding theory.}

\as{The key objects in this theory are the monopole charge operators $Q_{\vb*{r}}$, which in the exact theory are equal to $\eta_{\vb*{r}}\sum_{j \in \vb*{r}} \sigma^z_j$, located on each tetrahedron site $\vb*{r}$. The factor of $\eta_{\vb*{r}}$, taking value $1$ ($-1$) if the tetrahedron is on the up-pointing (down-pointing) sublattice, ensures that a spinon hopping between different tetrahedron sublattices retains the same charge.
In GMFT, these charges $Q_{\vb*{r}}$ are replaced by angular-momentum operators canonically conjugate to quantum rotors $\phi_{\vb*{r}}$, extending their spectrum from the phyisically allowed $\{-2,-1,0,1,2\}$ to all of $\mathbb{Z}$. The $\phi_{\vb*{r}}$ and $\phi_{\vb*{r}}^*$ operators then raise and lower the spinon number on a given site by $1$. Notice that, 
upon extending to $\mathbb{Z}$, 
the operators do not annihilate any state.}

\as{
After 
\begin{enumerate}
    \item integrating out $Q_{\vb*{r}}$ fluctuations, 
    \item replacing the fluctuating photon field $a$ by a mean-field expectation value, and
    \item introducing a Lagrange multiplier $\lambda_{{\vb*{r}},\tau}$ to enforce the unit-length rotor constraint $\phi^*_{{\vb*{r}},\tau} \phi_{{\vb*{r}}\tau} = 1$,
\end{enumerate}
one obtains the GMFT action
\begin{align}
S_{\rm GMFT} &= \int_0^{\beta} d\tau~  \mathcal{H}_{\rm rotor}[\phi, \phi^*] \nonumber \\
			 & +\sum_{\vb*{r}} \frac{\partial_\tau \phi^*_{\vb*{r}\tau}\partial_\tau \phi_{\vb*{r}\tau}}{2 J_{zz}} + i\lambda_{\vb*{r}\tau} (\phi^*_{\vb*{r}\tau} \phi_{\vb*{r}\tau} -1) \label{eq:quadratic_action}
    \, , \\
\mathcal{Z} &= \int \mathcal{D}[\phi^*, \phi, \lambda]e^{-S_{\rm GMFT}}
\, . 
\end{align}
Here, $\mathcal{H}_{\rm rotor}$ is simply $\mathcal{H}'$ in the main text, with the spin ladder operators replaced by their spinon representation $\sigma^{+}_{r_{\wedge}+\vb*{e}_\alpha/2} \mapsto \frac{1}{2}\phi^{*}_{\vb*{r}_\wedge}e^{ia_{\vb*{r}_{\wedge},\vb*{r}_{\wedge}+\vb*{e}_{\alpha}}} \phi_{\vb*{r}_\wedge+\vb*{e}_\alpha}$, as shown below for completeness:
\begin{align*}
	\mathcal{H}_{\rm rotor} &= 
	- \frac{J_{\pm}}{4}\sum_{\vb*{r} }\sum_{\alpha \neq \beta} \phi^{*}_{\vb*{r}+\eta_{\vb*{r}}\vb*{e}_\alpha}\phi_{\vb*{r}+\eta_{\vb*{r}}\vb*{e}_\beta}e^{i\eta_{\vb*{r}}(a_{\vb*{r},\vb*{r}+\vb*{e}_\beta}-a_{\vb*{r},\vb*{r}+\vb*{e}_\alpha})} \\
	+& \sum_{\vb*{r}_\wedge} \sum_{\alpha=0}^3\frac{\hat{\vb*{z}}_{\alpha}\cdot \vb*{B}}{4}\left[\phi^{*}_{\vb*{r}_\wedge}e^{ia_{\vb*{r}_\wedge,\vb*{r}_\wedge+\vb*{e}_{\alpha}}} \phi_{\vb*{r}_\wedge+\vb*{e}_\alpha}
+ H.c.  \right]
\, . 
\end{align*}
As in the main text, $\alpha,\beta \in \{0,1,2,3\}$ are spin sublattice indices. The sum over $\vb*{r}$ runs over all tetrahedra, while that over $\vb*{r}_\wedge$ only includes up-pointing tetrahedra. The two terms in this spinon Haimltonian are isotropic second-neighbour hopping, and field-tunable anisotropic nearest-neighbour hopping, respectively.}

\as{
We then make a large-$N$ approximation that relaxes the $\phi^* \phi = 1$ constraint to $\langle \phi^* \phi \rangle = 1$, replacing the field $\lambda_{r\tau}$ by a constant $\lambda$ that acts as an adjustable chemical potential for the spinons. Values for $\lambda$ can be obtained self-consistently by enforcing the $\langle \phi^*\phi \rangle =1$ constraint. This step renders Eq.~\eqref{eq:quadratic_action} quadratic, and therefore soluble by diagonalising the Hamiltonian in momentum space. Much like the case of electrons coupled to a magnetic field, the spinon dispersion relation can be shifted in momentum space by gauge transformations.}

\as{
Finally, we compute the dynamical magnetic structure factor $\mathcal{S}(\mathbf{q}, \omega) =\left( \delta^{\alpha \beta} - \frac{q^\alpha q^\beta}{|q|^2} \right) \langle m^\alpha(q, \omega) m^\beta(-q, 0) \rangle$, where the onsite magnetisation vectors are defined (in real space) by $\vb*{m}(\vb*{r}_\wedge+\vb*{e}_\rho/2) = \vb*{\hat{z}}_\rho (\tilde{g}_{zz}S^z + \tilde{g}_{xz}S^x) = \vb*{\hat{z}}_\rho \sigma^x(\vb*{r}_\wedge+\vb*{e}_\rho/2)$ since we take $\tilde{g}_{xz} = 0$. This amounts to evaluating the four-spinon correlator $\langle \phi^* \phi^* \phi \phi \rangle$, which is independent of the emergent gauge field.
For details of this calculation, see Ref.~\onlinecite{desrochersSymmetryFractionalizationGauge2023}.}

\as{
For the purposes of carrying out explicit calculations, it is necessary to find lattice gauge field configurations yielding the required flux background. Like the $U(1)_0$ and $U(1)_\pi$ cases, a simple analytic Ansatz (consisting of, respectively, one and four primitive unit cells) for the emergent mean-field vector potential $a$ is capable of generating the $\pi\pi00$ flux configuration. However, the situation is more computationally difficult in the frustrated case. The unit cell needed to contain a minimal $a$ configuration is generally very large: within a periodic unit cell, all plaquette fluxes of a particular sublattice must sum to an integer multiple of $2\pi$. Fluxes which are irrational multiples of $2\pi$ correspond to incommensurate gauge field configurations. We generate a [111]-field FF state capturing a flux $3\Phi$ through the [111]-facing plaquette and $-\Phi$ through the other three plaquette sublattices, using the explicit construction 
\begin{align}
    a_{\vb*{r}_{\wedge,0}+n_i\vb*{a}_i + \vb*{e}_\rho/2}
    = 
    n_i\mathbf{M}_{i\rho} \\
    \mathbf{M} = \begin{pmatrix}
        0 & 0 & -\Phi & -\Phi \\ 0 & 0 & 0 & 0\\ 0 & 0 & \Phi & 0
    \end{pmatrix}
    \, , 
    \label{eq:landau_gauge}
\end{align}
where $\vb*{a}_{x,y,z}$ are the three primitive lattice vectors and $\vb*{r}_{\wedge,0}$ is an arbitrary reference up-pointing tetrahedron. Such a construction becomes commensurate with the unit cell when $\Phi = q\pi/2$ for rational $q$.}

\as{
In Fig.~\ref{fig:grand-unified}, we present a comprehensive set of simulated correlations for the four flux phases of DO-QSI, intended mainly as a point of reference. Note that gauge fluctuations have been entirely neglected, and so the spinons shown here cannot exhibit any markers of the photon birefringence.
In both the 0-flux and $\pi$-flux phases, due to the generally flat dispersion, somewhat sharp peaks emerge in the two-spinon continuum. Counting such peaks once they have been split by the magnetic field allows for a direct experimental bound on the number of spinon bands. At a given point in $k$ space with $N$ spinon bands, there are $N(N+1)/2$ unique combinations of energies. Indeed, in Fig.~\ref{fig:field_piflux} the six sharp features seen imply $N\ge 3$. From the unit cell geometry we know $N$ to be even, and so can correctly deduce $N=4$. Even in the more dispersive $\pi\pi00$-flux phase, it is still possible to cleanly resolve $4$ peaks of spectral weight, ruling out $N=2$ and again correctly finding $N=4$. Finally, we remark on the frustrated-flux phase. We see a large number of closely-spaced, nearly-flat spinon bands; this band structure leads to a broad excitation in the two-spinon continuum seen by experiments.
}

\section{Direction dependence of field splitting in the $\pi$-flux phase}

\begin{figure}[h!]
	\includegraphics[width=\columnwidth]{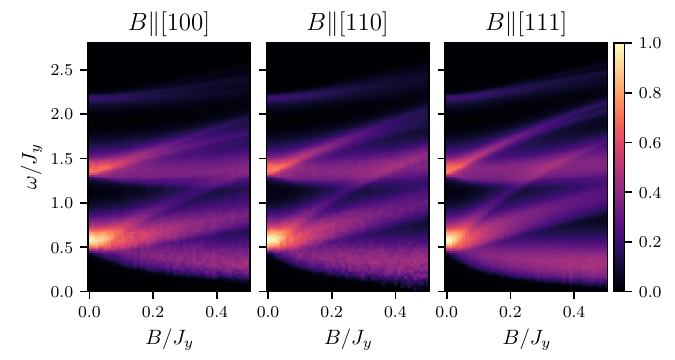}
	\caption{Weak direction-dependence of the spectral weight splitting in the $\pi$-flux phase. As in panels e) and f) of Fig.~\ref{fig:spinon-main}, the color scale reflects spectral weight integrated over the Brillouin zone.
	}
 \label{fig:field_piflux}
\end{figure}

\as{
In Fig.~\ref{fig:field_piflux}, we see that the spectral weight splittings under [100], [110] and [111] field are very similar. In this regime, the magnetic-field-mediated hopping is small compared to the $J_\pm$-mediated hopping and the $\pi$-flux phase is stable.
It follows that such signatures should be resolvable even in polycrystalline powder samples.}

\end{document}